\documentclass[showpacs,twocolumn,prd]{revtex4}

\usepackage{amssymb,amsmath} 
\usepackage{bbm} 
\usepackage{bm} 
\usepackage[english]{babel}

\usepackage{graphicx}
\usepackage{color}

\usepackage{float}

\usepackage[bookmarks,breaklinks,plainpages=false,pdfpagelabels]{hyperref}  
	\hypersetup{linkcolor=red,citecolor=blue,filecolor=dullmagenta,urlcolor=darkblue} 

\definecolor{grey}{rgb}{0.4,0.4,0.4}
\definecolor{dullmagenta}{rgb}{0.4,0,0.4}
\definecolor{darkblue}{rgb}{0,0,0.4}
\definecolor{orange}{rgb}{1,0.5,0}
\definecolor{lightbrown}{rgb}{0.75,0.5,0.25}
\definecolor{tan}{cmyk}{0.14,0.42,0.56,0}
\definecolor{djunglegreen}{cmyk}{0.99,0,0.52,0}
\definecolor{lightgreen}{rgb}{0,1,0}
\definecolor{olivegreen}{cmyk}{0.64,0,0.95,0.40}
\definecolor{midgreen}{rgb}{0.0,0.675,0.0}

\newcommand{\normsingle}[1]{\left|{#1}\right|}

\renewcommand{\a}{\ensuremath{\alpha}}

\newcommand{\e}{\ensuremath{\epsilon}}

\renewcommand{\l}{\ensuremath{\lambda}}
\newcommand{\s}{\ensuremath{\sigma}}

\newcommand{\G}{\ensuremath{\Gamma}}

\newcommand{\q}{\quad}
\newcommand{\qq}{\qquad}
\newcommand{\qqq}{\qquad\quad}

\newcommand{\hs}{\hspace}
\renewcommand{\.}{\hspace{0.5mm}}

\newcommand{\ra}{\ensuremath{\rightarrow}}

\newcommand{\lra}{\ensuremath{\leftrightarrow}}

\newcommand{\Arm}{\ensuremath{\mathrm{A}}}

\newcommand{\Grm}{\ensuremath{\mathrm{G}}}

\newcommand{\Jrm}{\ensuremath{\mathrm{J}}}

\newcommand{\Rrm}{\ensuremath{\mathrm{R}}}

\newcommand{\Urm}{\ensuremath{\mathrm{U}}}
\newcommand{\Vrm}{\ensuremath{\mathrm{V}}}
\newcommand{\Wrm}{\ensuremath{\mathrm{W}}}

\newcommand{\Yrm}{\ensuremath{\mathrm{Y}}}

\newcommand{\arm}{\ensuremath{\mathrm{a}}}

\newcommand{\hrm}{\ensuremath{\mathrm{h}}}
\newcommand{\irm}{\ensuremath{\mathrm{i}}}

\newcommand{\prm}{\ensuremath{\mathrm{p}}}

\newcommand{\Fcal}{\ensuremath{\mathcal{F}}}

\newcommand{\Lcal}{\ensuremath{\mathcal{L}}}

\newcommand{\Ocal}{\ensuremath{\mathcal{O}}}
\newcommand{\Pcal}{\ensuremath{\mathcal{P}}}

\newcommand{\Scal}{\ensuremath{\mathcal{S}}}

\newcommand{\Zcal}{\ensuremath{\mathcal{Z}}}

\newcommand{\Isf}{\ensuremath{\mathsf{I}}}

\newcommand{\Nsf}{\ensuremath{\mathsf{N}}}

\newcommand{\Rbb}{\ensuremath{\mathbb{R}}}

\newcommand{\Jbbm}{\ensuremath{\mathbbm{J}}}

\newcommand{\onebbm}{\ensuremath{\mathbbm{1}}}

\newcommand{\kbm}{\ensuremath{\bm{k}}}

\newcommand{\pbm}{\ensuremath{\bm{p}}}

\newcommand{\xbm}{\ensuremath{\bm{x}}}
\newcommand{\ybm}{\ensuremath{\bm{y}}}

\newcommand{\D}{\ensuremath{\mathcal{D}}}
\renewcommand{\d}{\ensuremath{\mathrm{d}}}
\newcommand{\ee}{\ensuremath{\mathrm{e}}}
\newcommand{\sumint}[1]{\underset{#1}{\sum\hs{-4.6mm}\int\hs{0.8mm}}}
\newcommand{\defas}{\mathrel{\mathop :}=} 
\newcommand{\hph}[1]{\hphantom{#1\;\,}}
\newcommand{\Tr}{\ensuremath{\mathrm{Tr}}}

\newcommand{\Vol}{\ensuremath{\mathrm{Vol}}}

\renewcommand{\t}{\text}
\newcommand{\ub}[3]{\underbrace{#1}_{\phantom{#3}\,#2\,#3}}

\renewcommand{\ol}{\overline}

\newcommand{\os}{\overset}

\newcommand{\eg}{e.g.}
\newcommand{\ie}{i.e.}

\newcommand{\hc}{{\rm H.c.}}
\newcommand{\rhs}{r.h.s.}
\newcommand{\lhs}{l.h.s.}
\newcommand{\wrt}{w.r.t.}
\newcommand{\cf}{c.f.}
\newcommand{\cds}{\,\cdot\,}

\begin{document}

\title{Stochastic Inflation and Replica Field Theory}

\author{Florian K{\"u}hnel}
\email{kuehnel@physik.uni-bielefeld.de}
\author{Dominik J. Schwarz}
\email{dschwarz@physik.uni-bielefeld.de}
\affiliation{Fakult{\"a}t f{\"u}r Physik, Universit{\"a}t Bielefeld, Postfach 100131, 33501 Bielefeld, Germany}
\date{\today}

\begin{abstract}
We adopt methods from statistical field theory to stochastic inflation. For the example of a free test field in de Sitter and power-law inflation, the power spectrum of long-wavelength fluctuations is computed. We study its dependence on the shape of the filter that separates long from short wavelength modes. While for filters with infinite support the phenomenon of dimensional reductions is found on large super-horizon scales, filters with compact support return a scale-invariant power spectrum in the infra-red. 
Features of the power spectrum, induced by the filter, decay within a few $e$-foldings. Thus the late-time power spectrum is independent of the filter details.
\end{abstract}


\maketitle

\section{Introduction}
\label{sec:Introduction}
Despite of its successes as a building block of our current picture of the Universe, a full understanding of the inflationary dynamics on super-horizon scales is still lacking. In his pioneering work \cite{1982PhLB..117..175S}, {Starobinsky} introduced the concept of {stochastic inflation} to provide a framework to study the evolution of quantum fields in an inflationary universe \cite{1986LNP...246..107S,PhysRevD.27.2848,Goncharov:1987ir,1987NuPhB.284..706R,1989PhRvD..39.2245K,linde-1994-49}. The key idea, which acquired considerable interest over the last years \cite{Bellini:2000rb,Bellini:2000er,Kunze:2003vp,liguori-2004-0408,Hattori:2005ac,PhysRevD.61.084008,2005PhRvD..71f3514M,MadrizAguilar:2006cw,martin-2006-73,martin-2006-73-a,Breuer:2006cd,Kunze:2006tu,Li:2007uc,Tolley:2008na}, lies in splitting the quantum fields into long- and short-wavelength modes, and viewing the former as classical objects evolving stochastically in an environment provided by quantum fluctuations of shorter wavelengths. Hence, it constitutes an example of how fundamental properties of quantum fields can be modeled using methods of statistical mechanics.

Given the de Sitter horizon, $\chi$, as a natural length scale of the problem, one then focusses on the ``relevant'' degrees of freedom (the long-wavelength modes) and regards the short-wavelength modes as ``irrelevant'' ones, where ``short'' and ``long'' are subject to $\chi$.

The most simple setup provides a fixed cosmological background, in which the dynamics of a scalar test field $\varphi$ is analyzed. If $\varphi$ is free, massive and minimally coupled, one obtains after splitting into long and short wavelengths, $\varphi = \varphi_{\t{\tiny L}} + \varphi_{\t{\tiny S}}$, an effective equation of motion of generalized Langevin-type,
\begin{align}
	\big( \Box + \mu^{2} \big) \varphi_{\t{\tiny L}}( t, \xbm )
		&=						\hrm( t, \xbm ) .
								\label{eq:linearfieldeq}
\end{align}
In equation \eqref{eq:linearfieldeq}, $\varphi_{\t{\tiny L}}$ is viewed as a classical entity, evolving stochastically in the presence of a (quantum) random force $\hrm$, which is {Gaussian} distributed with zero mean.

Self-interactions cause deviations from the simple Langevin-type form, manifested in higher powers of $\varphi_{\t{\tiny L}}$ on both sides of Eq.~\eqref{eq:linearfieldeq} with coefficients being of stochastic origin. The methods presented in this work allow to analyze this most general test-field case, which, to our knowledge, has not been addressed so far. Early studies focussed on homogeneous fields, thus restricting attention to the {time} evolution of $\varphi_{\t{\tiny L}}$. Recently, we presented a method to calculate arbitrary $2$-point functions for general stochastic potentials \cite{kuhnel-2008}.

This work extends our previous results \cite{kuhnel-2008} and is devoted to the study of the scaling behavior and time evolution of the power spectrum of $\varphi_{\t{\tiny L}}$ in a fixed background. This is done by means of replica field theory, which is well known in statistical physics \cite{1991JPhy1...1..809M}. Replica field theory allows us to compute the non-coincident two-point function of the long-wavelength modes for the most general test-field case. In order to distinguish short from long wavelength, a filter is introduced. The most common filter in stochastic inflation is a (sharp) step function. We study in detail the dependence of the power spectrum on the shape and parameters of smooth filter functions and consider the sharp step as a limiting case.
For filter functions with infinite support, a variant of the so-called dimensional reduction \cite{efetov-larkin-1977,PhysRevLett.37.1364,PhysRevLett.43.744,1977JPhC...10L.257Y,1984CMaPh..94..459K,doi:10.1088/0305-4470/21/22/011} is found, which results in a strong deviation from scale invariance in the infra-red, signalling a breakdown of the test-field assumption \cite{kuhnel-2008}. However, we show in this work that scale invariance is preserved on all scales at late times for filters with compact support.

Our work is organized as follows: After calculating the free, noiseless propagator for exponential as well as for power-law inflation (Sec.~\ref{sec:Free_Power_Spectrum}), we add quantum noise in Sec.~\ref{sec:Adding_Noise}. Its effect on the power spectrum is discussed in Sec.~\ref{sec:Application_to_Cosmological_Observables}. Special focus is put on the dependence on the filter functions and a non-linearity parameter $g_{\t{\tiny NL}}$ is introduced to quantify the modification of the power spectrum. The methods needed, are described in appendices \ref{sec:Noise_Distributions_and_Replica_Trick} and \ref{sec:Variational_Calculation}. In particular, the replica trick is introduced and a variational technique is presented. We conclude with a summary.

\section{Free Power Spectrum}
\label{sec:Free_Power_Spectrum}
This section is devoted to review the quantization of a free, minimally coupled, $N$-component, real test field $\vec{\varphi}$ with mass $\mu$. We solve the classical mode equations, construct the propagator, the power spectrum and give expressions for the spectral index.

Let us concentrate on a spatially-flat, isotropic and homogeneous universe in four-dimensional space-time. For its scale factor we assume either {\it exponential inflation}, $a( t ) \defas \ee^{H t}$, with {\it Hubble rate} $H$, or {\it power-law inflation}, $a( t ) \defas ( t / t_{1} )^{p}$ with $p > 1$ and the reference time $t_{1}$ defined by $a_{1} \defas a( t_{1} ) = 1$. For convenience use $\hslash = c = 1$ and set either $H = 1$ or $t_{1} = 1$, respectively. The mode function $u( t, k)$ is defined via the decomposition of the field components ($i = 1, \ldots, N$)
\begin{align}
	\varphi_{i}\!\left( t, \kbm \right)
		&=						\hat{\arm}_{i}( \kbm ) u( t, k ) + \hc,
								\label{eq:varphi-fourier-decomposition}
\end{align}
with the modulus of the comoving momentum $k \defas \normsingle{\kbm}$. The annihilation and creation operators obey the commutation relations
\begin{subequations}
\begin{align}
	\big[ \hat{\arm}_{i}( \kbm ), \hat{\arm}_{j}^{\dagger}( \pbm ) \big]
		&=						( 2 \pi )^{3} \delta^{3}\!\left( \kbm - \pbm \right)\delta_{i j},\\
	\big[ \hat{\arm}_{i}( \kbm ), \hat{\arm}_{j}( \pbm ) \big]
		&=						0.
\end{align}
\end{subequations}

The rescaled mode functions $v\!\left( \tau, k \right) \defas a( \tau )\,u\!\left( \tau, k \right)$ fulfil the mode equation
\begin{align}
	v'' + \left[ k^{2} + \mu^{2}\,a^{2} - \frac{a''}{a} \right] v
		&=						0 ,
								\label{eq:modeequation-chi}
\end{align}
with primes denoting derivatives \wrt~{\it conformal time} $\tau$, defined by $\d \tau \defas a( t )^{-1} \d t$. Solutions to \eqref{eq:modeequation-chi} are fixed by requiring that for very short wavelengths the effect of space-time curvature and mass becomes irrelevant, and thus a plain-wave solution should be obtained, \ie,
\begin{align}
	\lim_{\frac{k}{a} \,\ra\, \infty} v\!\left( \tau, k \right)
		&=						\frac{\ee^{- \irm\,k\,\tau}}{\sqrt{2\,k}} .
								\label{eq:contraint-chi}
\end{align}
The factor $1 / \sqrt{2\,k}$ is fixed by the canonical commutation relations of $\varphi$ and its conjugate momentum.

Exponential inflation implies $a( \tau ) = \normsingle{\tau}^{-1}$ and for power-law models one finds $a( \tau ) \propto \normsingle{\tau}^{-\frac{p}{p - 1}}$, where both cases match in the limit $p \ra \infty$. Then, equation \eqref{eq:modeequation-chi} can be rewritten in the form
\begin{align}\delimitershortfall=-1pt
	v'' + \bigg[ k^{2} - \frac{1}{\tau^{2}}\Big( \nu^{2} - \frac{1}{4} \Big) \bigg] v
		&=						0 ,
								\label{eq:modeequation-chi-nu}
\end{align}
with
\begin{subequations}
\begin{alignat}{2}
	\qq\nu
		&=						\sqrt{\frac{9}{4} - \mu^2}					&\qqq\t{(exponential)},&
								\label{eq:nu-power-law}\displaybreak[1]
\intertext{and with $\mu = 0$}
	\qq\nu
		&=						\frac{p}{p - 1} + \frac{1}{2}					&\qqq\t{(power-law)}.&
								\label{eq:nu-exponential}
\end{alignat}
\end{subequations}

A general solution to \eqref{eq:modeequation-chi-nu}, fulfilling \eqref{eq:contraint-chi}, is given in terms of Bessel functions:
\begin{align}
	u\!\left( \tau, k \right)
		&=						\frac{\sqrt{\pi}}{2}\frac{\normsingle{\tau}^{1 / 2}}{a( \tau )}\big( \Jrm_{\nu}\!\left( k\normsingle{\tau} \right)
								+ \irm\,\Yrm_{\nu}\!\left( k\normsingle{\tau} \right) \!\big) .
								\label{eq:modeequation-chi-nu-solution}
\end{align}
On large scales (for $k \ra 0$) and for positive $\nu$ the leading term of \eqref{eq:modeequation-chi-nu-solution} is
\begin{subequations}
\begin{align}
	u\!\left( \tau, k \right)
		&\simeq					-\frac{\irm\,2^{\nu - 1}\normsingle{\tau}^{1 / 2 - \nu}\,\G( \nu )}{\sqrt{\pi}\,a( \tau )}\,k^{-\nu} ,
								\label{eq:modeequation-chi-nu-solution-small-k-positive-nu}
\end{align}
while for negative $\nu$ one has
\begin{align}
	u\!\left( \tau, k \right)
		&\simeq					\frac{2^{- \nu -1}\,\sqrt{\pi}\normsingle{\tau}^{1 / 2 + \nu}\,\big( \irm \cot( \pi\,\nu ) + 1 \big)}{\G( \nu + 1 )\,a( \tau )}\,k^{\nu} .
								\label{eq:modeequation-chi-nu-solution-small-k-negative-nu}
\end{align}
\end{subequations}
The {\it propagator} is defined as
\begin{align}
	\Grm_{0}( t, t'\!, \kbm, \kbm' )
		&\defas					\frac{1}{N} \big< \Omega \big| \vec{\varphi}( t, \kbm) \cdot \vec{\varphi}( t'\!, \kbm' ) \big| \Omega \big> ,
\end{align}
where the vacuum $\big| \Omega \big>$ is defined by $\hat{\arm}( k ) \big| \Omega \big> = 0$ and a subscript ``$0$'' indicates a quantity that is calculated in the absence of any classical noise. The factor $1 / N$ is convenience.

An object of central interest in cosmology is the dimensionless {\it power spectrum} $\Pcal_{\!\varphi}( k )$. Its relation to some field propagator $\Grm\!\left( k \right)$, with assumed infra-red behavior $\Grm\!\left( k \right) \sim k^{-\eta}$, is given by
\begin{align}
	\Pcal_{\!\varphi}( k )
		&\defas					k^{3} \Grm\!\left( k \right)
		\sim						k^{n_{\varphi} - 1},
								\label{eq:spectral-index}
\end{align}
with the {\it spectral index} $n_{\varphi}$. It is connected to the so-called {\it critical exponent} $\eta$ via $n_{\varphi} = 4 - \eta$.

For $\mu = 0$ the power spectrum of the free, noiseless theory is scale-invariant, \ie, $n_{\varphi} = 1$ for $\eta_{0} = 3$. Non-zero mass leads to
\begin{subequations}
\begin{align}
	n_{\varphi}
		&=						4 - 3 \sqrt{1 - \frac{ 4 }{ 9 } \mu^{2} }
		=						1 + \frac{ 2 }{ 3 } \mu^{2} + \Ocal\!\left( \mu^{4} \right)
								\label{eq:ns-exponential-free}
\end{align}
and the power-law case with $\mu = 0$ yields
\begin{align}
	n_{\varphi}
		&=						\frac{p - 3}{p - 1}
		=						1 - \frac{ 2 }{ p } + \Ocal\!\left( \frac{1}{p^{2}} \right) .
								\label{eq:ns-power-law-free}
\end{align}
\end{subequations}
Note, that $n_{\varphi}( \mu \ne 0 ) > 1$ for exponential inflation, while $n_{\varphi}( p < \infty ) < 1$ in the massless power-law case, but in the limit $\mu \ra 0$, or, $p \ra \infty$, respectively, one recovers a scale-invariance spectrum. Also note, that the results \eqref{eq:ns-exponential-free} and \eqref{eq:ns-power-law-free} do not include any metric perturbation, which would generally cause negative deviations from $n_{\varphi} = 1$ (see, \eg, \cite{PhysRevD.66.023515}).

\section{Adding Noise}
\label{sec:Adding_Noise}
Here, we describe the dynamics of the long-wavelength modes, $\vec{\varphi}_{\t{\tiny L}}$, of some quantum field $\vec{\varphi}$ and how they are influenced by its short-wavelength counterparts. In this sense, the notion of stochastic inflation refers to a stochastic modelling of the former (treated as classical) evolving in a (stochastic) environment provided by the latter.
Now, how does one obtain the aforementioned Langevin-type equation?

To demonstrate this idea, let us follow Kandrup \cite{1989PhRvD..39.2245K} \footnote{There seems to be an error in equation (3.7) of \cite{1989PhRvD..39.2245K} and in the conclusion drawn from it. For the correct formula we refer the reader to the work by Rey \cite{1987NuPhB.284..706R}.} and consider (in four dimensions and for $N = 1$) the equation of motion
\begin{align}
	\big(\Box + \mu^{2} \big)\varphi( t, \xbm )
		&=						0 .
								\label{eq:eom-adding-noise}
\end{align}
Then one splits the field $\varphi$ into a long and short-wavelengths part, $\varphi = \varphi_{\t{\tiny L}} + \varphi_{\t{\tiny S}}$, like 
\begin{align}
	\varphi_{\t{\tiny S}}( t, \xbm )
		&=						\int\d^{3} k\;\Theta\big( k\,\tau( t ) - \e \big)\times\notag\\
		&\hph{=}					\times\Big[ \hat{\arm}( \kbm )\,u( t, k )\,\ee^{- \irm\,\kbm\cdot\xbm}
								+ \hat{\arm}^{\dagger}\!( \kbm )\,u^{*}( t, k )\,\ee^{+ \irm\,\kbm\cdot\xbm} \Big] .
								\label{eq:stocastic-inflation-phi-lwl-a}
\end{align}
$\Theta$ is the usual step function that sharply cuts out the long wavelengths. Instead, one could also write an arbitrary, smooth filter function (\cf~the discussion in section \ref{sec:Filter_Functions}). The parameter $\e$, which basically says where to split into short and long wavelengths, should be chosen much smaller than one in order to cut far beyond the Hubble radius. On the other hand $\e$ is bounded from below, since otherwise, the change in the background metric has to be taken into account. This leads to $\e \gg \exp\{ - H^{2} / \dot H \}$, \ie, $\normsingle{\ln\{ \e \}} \gg 0$.

Inserting \eqref{eq:stocastic-inflation-phi-lwl-a} in \eqref{eq:eom-adding-noise} yields
\begin{align}
	\big( \Box + \mu^{2} \big) \varphi_{\t{\tiny L}}( t, \xbm )
		&=						\hrm( t, \xbm ) ,
								\label{eq:eom-stocastic-inflation-a}
\end{align}
with $\hrm$ being given by
\begin{align}
	\hrm( t, \xbm )
		&\defas					\int\d^{3} k\;\delta\big( k\,\tau( t ) - \e \big)\times\notag\\
		&\hs{-5mm}				\times\Big[ \hat{\arm}( k )\,w( t, \kbm, \xbm )\,\ee^{- \irm\,\kbm\cdot\xbm}
								+ \hat{\arm}^{\dagger}\!( k )\,w^{*}( t, \kbm, \xbm )\,\ee^{+ \irm\,\kbm\cdot\xbm} \Big] .
								\label{eq:stocastic-inflation-phi-lwl-2-a}
\end{align}
Here, $w( t, \kbm, \xbm )$ is obtained after the application of the covariant Laplacian (see \cite{1989PhRvD..39.2245K} for details). The point now is, that $\varphi_{\t{\tiny L}}$ is viewed as a classical entity (instead of as a quantum one) which evolves in the presence of the ``random'' force $\hrm$ that is itself treated as a genuine quantum object. This turns equation \eqref{eq:eom-stocastic-inflation-a} into a classical (generalized) Langevin equation with $\hrm$ being a {Gaussian} distributed random variable, with the moments (quantum averages) $\left\langle \hrm( t, \xbm ) \right\rangle = 0$ and $\left\langle \hrm( t, \xbm )\,\hrm( t', \xbm' ) \right\rangle \ne 0$.

For exponential inflation the noise two-point function is approximately given by \cite{1989PhRvD..39.2245K}
\begin{align}
	\big< \hrm( t, \xbm )\,\hrm( t', \xbm' ) \big>
		&\simeq					\frac{9}{4\,\pi^{2}}\,\delta( t - t' )\,\frac{\sin\!\Big( \frac{\e\,| \xbm - \xbm' |}{\tau( t )} \Big)}
								{\frac{\e\,| \xbm - \xbm' |}{\tau( t )}} .
								\label{eq:P(t,t',x-x')}
\end{align}
However, this result heavily relies on the particular choice of the filter function. The Markov property (manifest through the time delta function) is a consequence of the step function and does in general not occur for arbitrary smooth window functions. A more detailed discussion on these functions is presented in section \ref{sec:Filter_Functions}.

The Lagrangian $\Lcal\!\left( \varphi_{\t{\tiny L}} \right)$ for the long-modes $\varphi_{\t{\tiny L}}$ corresponding to equation \eqref{eq:eom-stocastic-inflation-a} has, despite of its free part $\Lcal_{0}$---associated with the \lhs~of \eqref{eq:eom-stocastic-inflation-a}---two additional pieces. The first provides the interaction with the noise and a second is quadratic in $\hrm$, which is ascribed to a stochastic probability distribution $\prm[ \hrm]$ [\cf~equation \eqref{eq:P[h]}]. The Lagrangian (including only $\varphi_{\t{\tiny L}}$) is
\begin{align}
	\Lcal\!\left( \varphi_{\t{\tiny L}} \right)
		&=						\Lcal_{0}\!\left( \varphi_{\t{\tiny L}} \right) + \hrm\,\varphi_{\t{\tiny L}}
								\label{eq:lagrange-a}
\end{align}
to which, in statistical field theory, is referred to as the {\it random-field} ({\sc rf}) case. As has been shown recently, this effectively re-sums the leading-log contribution of the full quantum theory \cite{Tsamis:2005hd}. Please note that equation \eqref{eq:lagrange-a} has been derived for free fields only.

As soon as interactions are taken into account it can (and in general will) be only an approximation. Since one is often interested in theories containing interaction potentials, and in addition in situations where quantum fluctuations may become large, a generic effective interacting theory will generate some stochastic potential $\Vrm$,
\begin{align}
	\Lcal\!\left( \vec{\varphi}_{\t{\tiny L}} \right)
		&=						\Lcal_{0}\!\left( \vec{\varphi}_{\t{\tiny L}} \right) + \Vrm\!\left( \vec{\varphi}_{\t{\tiny L}} \right) ,
								\label{eq:lagrange}
\end{align}
meaning that the Taylor coefficients of $\Vrm\!\left( \vec{\varphi}_{\t{\tiny L}} \right)$ are random variables (see below). Here and in the following, we generalize to $N$-component fields, with correlations $\Grm_{ij}( t, k ) \equiv \delta_{ij}\,\Grm( t, k )$. To simplify notation, we will drop the subscript `$\t{\footnotesize L}$' in the following.

In the random-field case the potential $\Vrm$ corresponds to $\Vrm_{\t{\tiny RF}}\!\left( \vec{\varphi}( x ) \right) = - \sum_{j}\hrm_{j}( x )\,\varphi_{j}( x )$, whereas the case of so-called {\it random anisotropy} ({\sc ra}) is described by $\Vrm_{\t{\tiny RA}}\!\left( \vec{\varphi}( x ) \right) = - \sum_{jk}\hrm_{jk}( x )\,\varphi_{j}( x )\,\varphi_{k}( x )$. In general one may has
\begin{align}
	\Vrm\big( \vec{\varphi}( x ) \big)
		&=						-\, \sum_{n=1}^{\infty}\sum_{i_{1} \dots i_{n}=1}^{N} \hrm_{i_{1}\ldots i_{n}}( x )\,
								\varphi_{i_{1}}( x )\cdot\ldots\cdot \varphi_{i_{n}}( x ) .
								\label{eq:general-stochastic-interaction-potential}
\end{align}
In all cases $\{ \hrm \}$ is assumed to be a set of {Gaussian}-distributed random variables which' distribution may be written as
\begin{align}
	\prm\big[ \{ \hrm \} \big]
		&\propto					\exp\!\Bigg\{ - \frac{1}{2}\sumint{\l, \l'}\hrm_{\l}\Arm_{\l, \l'}\hrm_{\l'} - \sumint{\l}\mu_{\l}\hrm_{\l} \Bigg\}
								\label{eq:P[h]}
\end{align}
with mean $\mu_{\l}$. To simplify notation further, we use a general index containing position and all other components, \ie, we write $\hrm_{\l}$ instead of $\hrm_{i_{1}\ldots i_{n}}( x )$.

Distribution \eqref{eq:P[h]} is an approximation for general interactions and justified for a limited class of theories only. However, for many systems it is at least a reasonable starting point, providing the possibility to obtain exact non-perturbative results. In this work we focus on the random-field noise, \ie, free, massive scalar test fields in various backgrounds.

For some quantity $\Ocal$ depending on $\{ \hrm \}$, the {\it stochastic average}, which shall be denoted by a bar, is then calculated as
\begin{align}
	\ol{\Ocal\left[ \{ \hrm \} \right]}	
		&\defas					\int{\!\D\left[ \{ \hrm \} \right]}\,\prm[ \{ \hrm \} ]\,\Ocal[ \{ \hrm \} ] .
\end{align}
Note, that linearizing the equation of motion in the quantum modes, corresponds {\it exactly} to a Gaussian distribution of mean zero. Therefore, \eqref{eq:P[h]} is not an approximation in this case.

The methods needed for the treatment of these stochastic systems are discussed in appendix \ref{sec:Noise_Distributions_and_Replica_Trick} and have been first introduced in the cosmological context in our recent work \cite{kuhnel-2008}.

\section{Effective Power Spectrum}
\label{sec:Application_to_Cosmological_Observables}
This section contains the study of the infra-red behavior of the physical propagator $\Grm$ and therefore of the power spectrum $\Pcal$, with special focus on the role of the filter.

\subsection{Long-Range Correlation}
\label{sec:Long-Range_Correlation}
The full two-point function for the long-wavelength modes, $\Grm \defas \frac{1}{N}\overline{\big< \vec{\varphi}_{\t{\tiny L}}\cdot \vec{\varphi}_{\t{\tiny L}} \big>}$, is calculated in appendix \ref{sec:Variational_Calculation}, using a Feynman-Jensen-type variational calculation together with replica field theory. The result is
\begin{align}
	\Grm(t, k)
		&=						\Grm_{0}(t, k) + \s(t, k)\,\Grm_{0}(t, k)^{2} ,
								\label{eq:G=G0-sigmaG02}
\end{align}
where the so-called replica structure $\s$ is determined from the variational equations \eqref{eq:sc+sab1N-times-lr} and \eqref{eq:sc+sab1N-difference-lr}, respectively.

To analyze the physical consequence of equation \eqref{eq:G=G0-sigmaG02} on the power spectrum, let us now assume for the two-point noise correlation $\overline{\hrm_{i}( \xbm )\,\hrm_{j}( \ybm )} = \phi( \xbm - \ybm ) \sim \normsingle{\xbm - \ybm}^{- 3 + \rho}$ with $\rho < 3$ (\cf~appendix \ref{sec:Long_Range_Correlation}). This is the case of so-called long-range correlation, introduced in \cite{kuhnel-2008}, and describes properly the infra-red limit of the physical model discussed below. In momentum space, the above choice implies $\phi( k ) \sim k^{- \rho}$ and hence $\s( t, k ) = \wp( t ) k^{- \rho}$ by virtue of \eqref{eq:sc+sab1N-times-lr}. For $\rho > - \eta_{0}$, the infra-red behavior of the power spectrum deviates from the noiseless result, and we find $\eta = 2 \eta_{0} + \rho$. This result is consistent with previous studies in flat space for a propagator with $\eta_{0} = 2$ \cite{PhysRevB.27.5875}.

What does this mean for the spectral index? Since $n_{\varphi} = 4 - \eta$, we find the result
\begin{align}
	n_{\varphi}
		&=						4 - 2 \eta_{0} - \rho
								\label{eq:nvarphi-rho}
\end{align}
if $\rho > - \eta_{0}$. Thus, one would find a dramatic change of the super-horizon power spectrum as compared to the case without noise. Concretely, for exponential inflation this implies a modification of the spectral index on large super-horizon scales if the spatial noise correlator decreases at most like $\normsingle{\xbm}^{- 6}$ if $\mu = 0$, while for finite mass this exponent changes to $- 6 + 2 / 3\,\mu^{2} + \Ocal( \mu^{4} )$. In the massless power-law model the power is given by $- 6 - 2 p^{-1} + \Ocal(p^{-2})$.

Equation \eqref{eq:nvarphi-rho} constitues one variant of the pheno-menon of {\it dimensional reduction} \cite{PhysRevLett.43.744,1984CMaPh..94..459K}, which can rigorously be proven to all orders in perturbation theory for $\rho = 0$ and for arbitrary non-random potentials (see especially \cite{1984CMaPh..94..459K} for a supersymmetric version of the proof). Because this effect originates from the second piece of $\Grm$ in \eqref{eq:G=G0-sigmaG02} it will be referred to as the {\it dimensional reduction part}.

Please note, that those changes only concern the power spectrum of the smoothed (classical) long-wavelength modes, which are influenced by their short (quantum) counterparts. It does not mean that the full quantum two-point function obeys dimensional reduction.
 
We should underline that the above statements on dimensional reduction do depend on the concrete choice of the filter function. Their influence on the power spectrum is discussed in section \ref{sec:Filter_Functions}.

A natural question is to ask on which scales does the effect of dimensional reduction show up. Let us therefore define the {\it transition scale} $k_{*}$ at which the two terms on the \rhs~of equation \eqref{eq:G=G0-sigmaG02} balance each other via
\begin{align}
	\Grm_{0}(t, k_{*})
		&\os{!}{=}					\s(t, k_{*})\,\Grm_{0}(t, k_{*})^{2} .
								\label{eq:kstar-def}
\end{align}
It separates two regions such that for $k \gg k_{*}$ the behavior is noiseless and for $k \ll k_{*}$, dimensional reduction holds.

\subsection{Stochastic Inflation}
\label{sec:Stochastic_Inflation}
Let us now return to our physical model of stochastic inflation. The split of the field $\vec{\varphi}$ into a long- and short-wavelength part, $\vec{\varphi} = \vec{\varphi}_{\t{\tiny L}} + \vec{\varphi}_{\t{\tiny S}}$, together with the free field equation, $( \Box + \mu^{2} )\vec{\varphi} = \vec{0}$, implies for the infra-red part of the propagator
\begin{align}
	\Grm( t, k )
		&\simeq					\Bigg| \left( \Box_{k} + \mu^{2} \right)\!\bigg[ \Wrm_{\!\kappa}\!\left( \frac{ k }{ a( t ) } - \e \right) u( t, k ) \bigg] \Bigg|^{2}
								\big| u( t, k ) \big|^{4}\!,
								\label{eq:G(t,k)}
\end{align}
where $\Box_{k}$ is the (spatially Fourier transformed) covariant Laplacian, $u( t, k )$ is the mode function from equation \eqref{eq:varphi-fourier-decomposition}, and $\Wrm_{\!\kappa}$ is a smooth high-pass filter (\cf~section \ref{sec:Filter_Functions}), cutting out the low frequencies below $\e$. The parameter $\kappa$ controls the width of the cut. In the limit $\kappa \ra 0$, $\Wrm_{\!\kappa}$ approaches a step function. Here we choose
\begin{align}
	\Wrm_{\!\kappa}( \cds )
		&=						\delimitershortfall=-1pt
								\frac{1}{\pi} \arctan\Big( \frac{\cds}{\kappa} \Big) + \frac{1}{2}
								\label{eq:filter-function-main}
\end{align}
and take $0 < \e \ll 1$ in order to separate at wavelength well below the Hubble rate $H (= 1)$, and $\kappa \ll \e$ to have a narrow transition region between quantum and classical modes. We do not impose any restriction on $\mu$ except that we demand the radicant in \eqref{eq:ns-exponential-free} to be positive, \ie, $\mu^{2} \le 9 / 4$.

Please not that the filter function \eqref{eq:filter-function-main} does not have a compact support, meaning that also modes from the far infra-red influence the quantum noise. A further discussion on filter functions, in particular of such with compact support, is presented in the subsequent section.

Using \eqref{eq:filter-function-main} the model given in \eqref{eq:G(t,k)} is of long-range-type (\cf~appendix \ref{sec:Noise_Distributions_and_Replica_Trick}) and implies
\begin{subequations}
\begin{align}
	\q \rho
		&=						3 \sqrt{1 - \frac{4}{9}\mu^2} - 2
		=						1 - \frac{2}{3} \mu^{2} + \Ocal\!\left( \mu^{4} \right)
								\label{eq:ns-exponential-free-RS-inflation}
\intertext{for exponential inflation, and}
	\q \rho
		&=						\frac{p+1}{p-1}	
		=						1+\frac{2}{p} + \Ocal\!\left( \frac{1}{p^{2}} \right)
								\label{eq:ns-exponential-free-RS-matter}
\end{align}
\end{subequations}
for massless power-law models. Thus for $k \ll k_{*}$ we obtain with \eqref{eq:nvarphi-rho}
\begin{subequations}
\begin{align}
	\q n_{\varphi}
		&=						6 - 9 \sqrt{ 1 - \frac{ 4 }{ 9 }\mu^2 }
		=						- 3 + 2 \mu^{2} + \Ocal\!\left( \mu^{4} \right)
								\label{eq:ns-exponential-free-RS-inflation}
\intertext{in the exponential case, and}
	\q n_{\varphi}
		&=						3\.\frac{p+1}{1 - p}
		=						-3-\frac{6}{p} + \Ocal\!\left( \frac{1}{p^{2}} \right)
								\label{eq:ns-exponential-free-RS-matter}
\end{align}
\end{subequations}
for power-law inflation with $\mu = 0$. We see that scale invariance of the effective power spectrum is destroyed even in the massless, exponential inflationary scenario in the infra-red limit. Also the power-law case changes drastically.

For scales $k \gg k_{*}$ the noiseless spectral index \eqref{eq:ns-exponential-free} is recovered. The late-time behavior of the transition scale $k_{*}$, defined in equation \eqref{eq:kstar-def}, can be calculated analytically:
\begin{widetext}
\begin{subequations}
\begin{align}\label{eq:kstar-exp}
	k_{*}
		&=						\left( \ee^{- t} \right)^{\frac{8-2 \sqrt{9-4 \mu^2}}{2 \sqrt{9-4 \mu^2}-2}} \pi^{-\frac{2}{\sqrt{9-4 \mu^2}-1}}
								\left( \frac{2^{2 \sqrt{9-4 \mu^2}-3} \left(5- 2\,\mu^2 - \sqrt{9-4 \mu^2} \right)
								\kappa^2 \G\!\left(\frac{1}{2} \sqrt{9 - 4\mu^2}\right)^4}{\left( \e^2+\kappa^2\right)^2}\right)^{\frac{1}{2 \sqrt{9-4 \mu^2}-2}}
\end{align}
for exponential inflation and
\begin{align}\label{eq:kstar-power}
	k_{*}
		&=						\delimitershortfall=-1pt \Big( \frac{p}{t^{p + 1}} \Big)^{\frac{1}{2} - \frac{1}{2p}}\,2
								\left( \frac{\kappa\,\G\!\left(\frac{3}{2} + \frac{1}{p - 1} \right)^2}
								{\pi^{2} \left( \e^2 + \kappa^2 \right)} \right)^{\frac{1}{2}-\frac{1}{2p}}
\end{align}
\end{subequations}
\end{widetext}
for massless power-law inflation. In the zero-mass limit \eqref{eq:kstar-exp} yields the asymptotic form
\begin{align}
	k_{*}( t )
		&=						\ee^{- t / 2} \frac{2 \sqrt{\kappa}}{\sqrt{\pi}\sqrt{\e^2 + \kappa^2}}.
								\label{eq:kstar-exp-mu=0}
\end{align}
Thus, for $\e \ne 0$, $k_{*}$ goes to zero in the (step-function) limit $\kappa \ra 0$, \ie, dimensional reduction is absent---a statement that holds for exponential and power-law inflation as well. This is a general feature of the free (Gaussian) theory where is no mixing of the short quantum modes with the long classical ones as a sharp cut-off is introduced. In a slightly different setup, with a filter function having only one parameter, \ie, $\e = \kappa$, this has already been noted in reference \cite{matarrese-2004-0405}.

Let us turn to the issue of the compatibility of \eqref{eq:kstar-power} with \eqref{eq:kstar-exp}. As has already been mentioned in section \ref{sec:Free_Power_Spectrum}, for $p \ra \infty$ and $\mu = 0$, both cases should match. A naive limit of \eqref{eq:kstar-power} shows that this is not obvious. The point here is that one should carefully look at the time dependence. Expressing \eqref{eq:kstar-power} in terms of the number of $e$-foldings, $\Nsf \defas \ln\{ a / a_{1} \}$, shows indeed the desired coincidence. So for all plots related to power-law cases, we will make the replacement
\begin{align}
	t
		&=						\Big( p\,\ee^{\tilde{t}} \Big)^{\frac{1}{p + 1}} .
								\label{eq:modifed-time}
\end{align}
A straightforward calculation shows the relation between $\tilde{t}$ and $\Nsf$,
\begin{align}
	\Nsf
		&=						\frac{p}{p + 1}\big[ \tilde{t} + \ln\{ p \} \big]
		=						\tilde{t} + \ln\{ p \} + \Ocal\big( p^{-1} \big) .
\end{align}
This means that for large $p$, $\tilde{t} = \Nsf$ up to a shift which originates from different time normalization of the power-law and the de Sitter case where $t = \Nsf$. This shift is indeed visible in figures \ref{fig:kstar-power--law-expansion---modified-time-dependence} and \ref{fig:compare-filter-functions---power-law---modified-time-dependence}, where only the last part of the transient phenomenon does show up, contrary to the corresponding exponential inflation plots, where a larger part can be observed (details below).

\begin{figure}
	\centering
	\includegraphics[angle=0,scale=1]{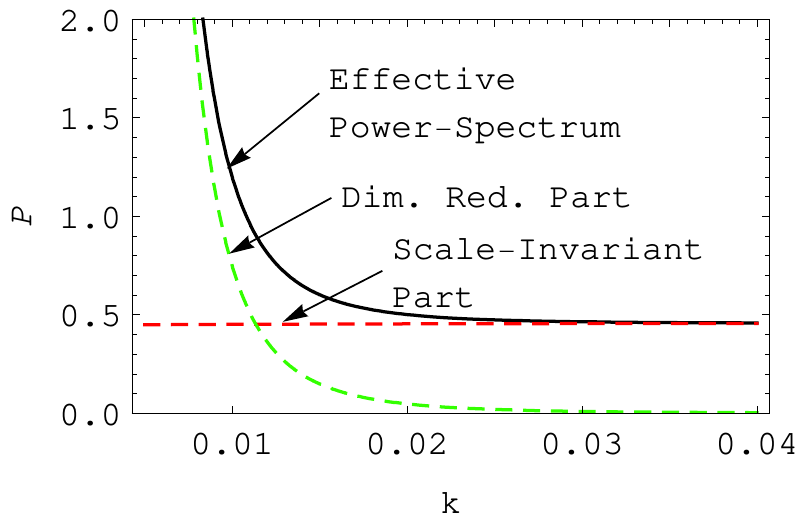}
	\caption{Power spectrum $\Pcal$ of a massive test field ($\mu = 0.1 H$) for exponential inflation at $t H = 10$ as a function of comoving momentum $k$. 
			The parameters of the filter function \eqref{eq:filter-function-main} are fixed to $\kappa = 10^{-3}$ and $\e = 10^{-2}$.}
	\label{fig:Power-spectrum-(exponential)}
\end{figure}

\begin{figure}
	\centering
	\includegraphics[angle=0,scale=1]{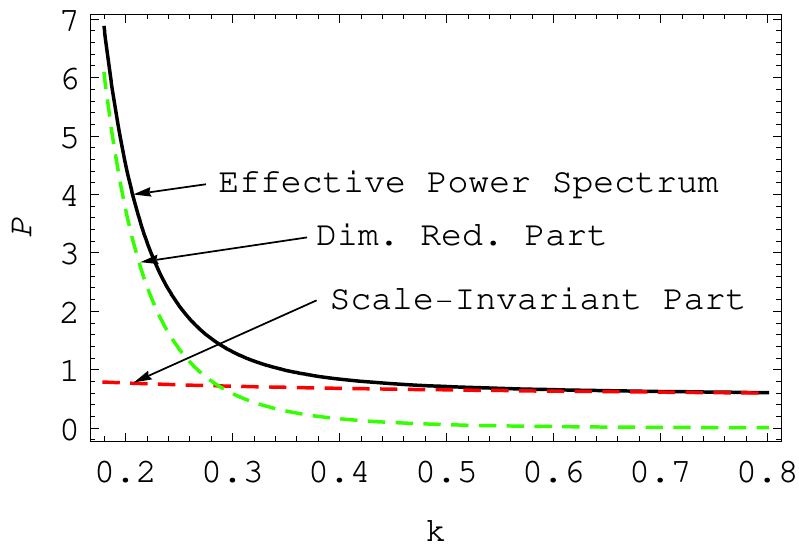}
	\caption{Power spectrum $\Pcal$ for a massless test field for power-law inflation ($p = 12$) at $\tilde{t} = 4$ [\cf~\eqref{eq:modifed-time}] as a function of comoving 
			momentum $k$ (in units of $1 / t_{1}$). The parameters of the filter function \eqref{eq:filter-function-main} are fixed as in figure 
			\ref{fig:Power-spectrum-(exponential)}.}
	\label{fig:Plots-of-P-(power-law)--modified-time-dependence-t=4}
\end{figure}

\begin{figure}
	\centering
	\includegraphics[angle=0,scale=1]{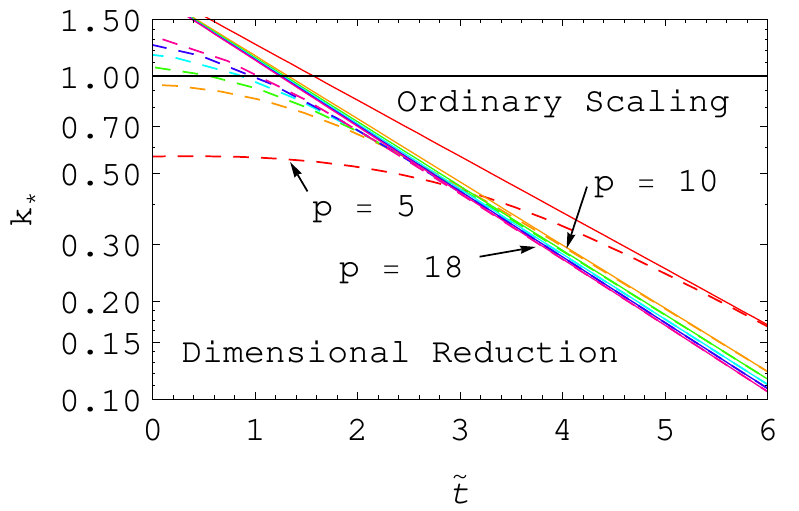}
	\caption{Comoving transition scale $k_{*}$ for power-law inflation as a function of modified cosmic time $\tilde{t}$ [\cf~\eqref{eq:modifed-time}] for 
			$p = 5, 10, 12, 14, 16, 18\t{ (dashed lines, top to bottom)}$. Dashed curves are numerical results, 
			colored solid lines are analytic approximations. The parameters of the filter function \eqref{eq:filter-function-main} are fixed as in 
			figure \ref{fig:Power-spectrum-(exponential)}.}
	\label{fig:kstar-power--law-expansion---modified-time-dependence}
\end{figure}

In figures \ref{fig:Power-spectrum-(exponential)} (and \ref{fig:Plots-of-P-(power-law)--modified-time-dependence-t=4}) we show the effective long-wavelength power spectrum $\Pcal$ as a function of $k$ for fixed time $H t = 10$ and mass $\mu = 0.1$ ($\tilde{t} = 4$ and $p = 12$) with $\e = 10^{-2}$ and $\kappa = 10^{-3}$ for the de Sitter (massless power-law) model. One can see that it diverges stronger than the noiseless power spectrum as $k$ tends to zero, putting therefore more correlation on large scales. One further sees that the part $k^{3} \s \Grm_{0}^{2}$ approximates the function the full power spectrum $\Pcal$ in the infra-red as well as that the noiseless piece $k^{3} \Grm_{0}$ gives a suitable ultra-violett approximation. In all the plots we took the filter function \eqref{eq:filter-1}.


\begin{figure}
	\centering
	\includegraphics[angle=0,scale=1]{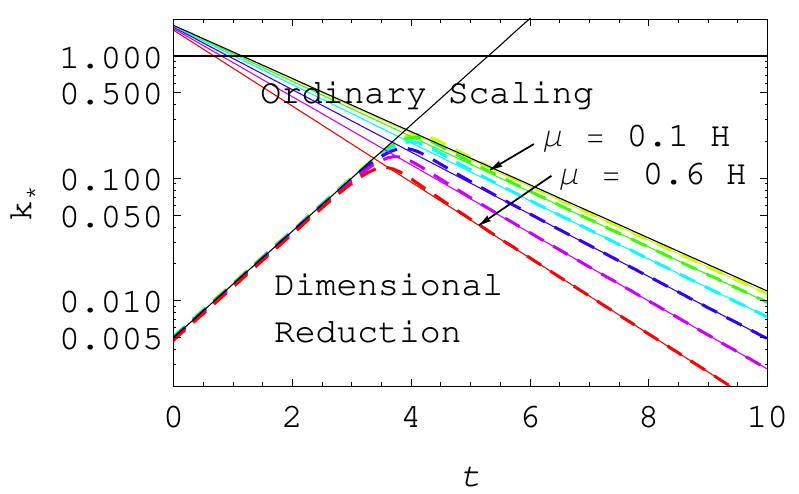}
	\caption{Comoving transition scale $k_{*}$ [\cf~\eqref{eq:kstar-exp}] for exponential inflation as a function of cosmic time $t$ (in units of $H$) 
			for mass $\mu / H = 0.1, 0.2, 0.3, 0.4, 0.5, 0.6\t{ (dashed lines, top to bottom)}$. Dash-ed curves are numerical results, 
			colored solid lines are analytic approximations, 
			and enveloping black lines are $\frac{ \e }{ 2 } a( t )$ and the asymptotic form \eqref{eq:kstar-exp-mu=0}, 
			respectively. The parameters of the filter function \eqref{eq:filter-function-main} are fixed as in figure \ref{fig:Power-spectrum-(exponential)}.}
	\label{fig:kstar-exponential-correlation}
\end{figure}

Figure \ref{fig:kstar-exponential-correlation} shows the time behavior of the comoving $k_{*}$ for exponential inflation for different values of the mass $\mu$. Figure \ref{fig:kstar-power--law-expansion---modified-time-dependence} displays the same for the power-law model. The solid rays represent the analytic approximations \eqref{eq:kstar-exp} and \eqref{eq:kstar-power}, respectively, while the dashed curves are the full results, obtained numerically from \eqref{eq:kstar-def} using the procedure \texttt{FindRoot} of {\sc Mathematica} 6. Well below this borderline the two-point function obeys dimensional reduction, while well above ordinary scaling holds. After an initial transient phenomenon, which' duration depends on the specific choice of $\e$ and $\kappa$, the comoving transition scale decays exponentially fast. Hence, the dimensional-reduction contribution is pushed to larger and larger scales as time increases. This therefore guarantees that quantum modes induce only a minor change of the spectral index on sub-horizon as well as on moderate super-horizon scales.

For concreteness, let us consider a mode with comoving $k = 0.05\.H$. At time $t = 0$ it is within the region of ordinary scaling, suffering at most slightly from dimensional reduction. This mode enters then, after roughly two $e$-foldings, the region of broken scale invariance, but leaves it at the latest (for $\mu = 0$) after seven $e$-foldings and stays eternally in the scale-invariant regime, which itself grows exponentially fast.


\subsection{Filter Functions}
\label{sec:Filter_Functions}
The discussion of possible smooth filter functions and their influence on the phenomenon of dimensional reduction shall be subject of this subsection. In particular, we study their effects on the transition scale $k_{*}( t )$.

Let $\Theta$ be the Heaviside function. Then, we define a {\it filter function} as a function $\Wrm_{\!\kappa}$ depending on a parameter $\kappa$ (controlling the width of the transition) such that
\begin{align}
	\lim_{\kappa \ra 0} \Wrm_{\!\kappa}
		&\equiv					\Theta .
								\label{eq:filter-function-def}
\end{align}

One may divide filter functions fulfilling \eqref{eq:filter-function-def} into two classes: Those for which $\Theta - \Wrm_{\!\kappa}$ has an infinite support $\Isf$ and those for which $\Isf$ is compact. Let us now discuss those two cases separately.

\subsubsection{Non-Compact Support}
\label{sec:Non-Compact_Support}
Some well-known smooth filter functions are:
\begin{subequations}
\begin{align}
	&\frac{\tan^{-1}\!\left( \frac{x}{\kappa} \right)}{\pi} + \frac{1}{2} ,
								\label{eq:filter-1}\displaybreak[1]\\
	&\frac{1}{2}\,{\rm erf}\!\left( \frac{x}{\kappa} \right) + \frac{1}{2} ,
								\label{eq:filter-3}\displaybreak[1]\\
	&\frac{{\rm Si}\!\left(\frac{\pi\,x}{\kappa} \right)}{\pi} + \frac{1}{2} ,
								\label{eq:filter-4}\displaybreak[1]\\
	&\frac{1}{1+\ee^{-\frac{x}{\kappa}}} ,
								\label{eq:filter-5}\displaybreak[1]\\
	&\ee^{-\ee^{-\frac{x}{\kappa}}} ,
								\label{eq:filter-6}\displaybreak[1]\\
	&\frac{1}{2}\tanh\!\left( \frac{x}{\kappa} \right) + \frac{1}{2} .
								\label{eq:filter-7}
\end{align}
\end{subequations}
Here, ${\rm erf}$ is the error function, defined by ${\rm erf}( z ) \defas \frac{2}{\sqrt{\pi}}\int_{0}^{z}\d t\;\ee^{-t^{2}}$ and ${\rm Si}$ is the sine-integral function, defined by ${\rm Si}( z ) \defas \int_{0}^{z}\d t\,\sin( t ) / t$.

The following two figures show the dependence of the transition scale on various filter functions, where we choose the functions \eqref{eq:filter-1}, \eqref{eq:filter-5} and \eqref{eq:filter-7}. It can clearly be seen that only the quantitative behavior changes, \ie, the position of the bump, which marks the end of the transient phenomenon, and {\it not} the qualitative shape. Note, that the curves are rescaled by a fixed factor (one for each filter function) and that those functions all have precisely the same asymptotic behavior. These statements concern both the exponential as well as the power-law case.
\begin{figure}
	\centering
	\includegraphics[angle=0,scale=1]{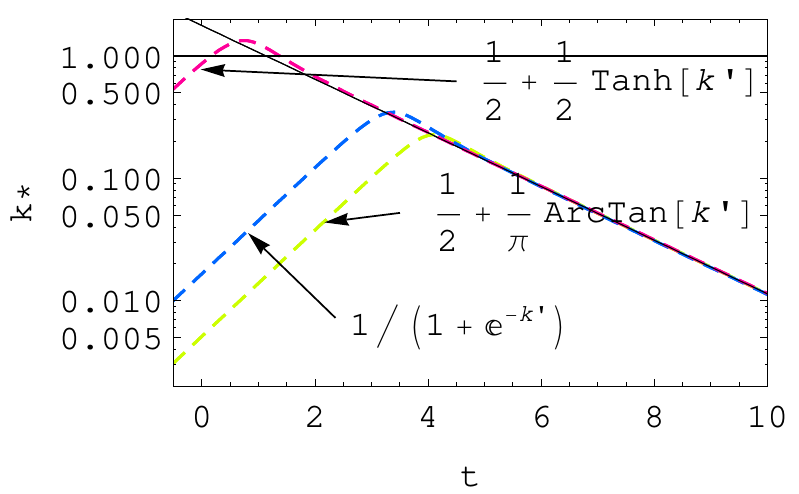}
	\caption{Influence of filter function shape on the comoving transition scale $k_{*}$ for exponential inflation as a function of cosmic time $t$ (in units of $H$) 
			for mass $\mu / H = 0.1$. A smoothing $\kappa = 10^{-3}$ and short-wavelength cut $\e = 10^{-2}$ are chosen. 
			The variable $k'$ is a short-hand notation for $( k \tau - \e ) / \kappa$.}
	\label{fig:compare-filter-functions---exponential}
\end{figure}
\begin{figure}
	\centering
	\includegraphics[angle=0,scale=1]{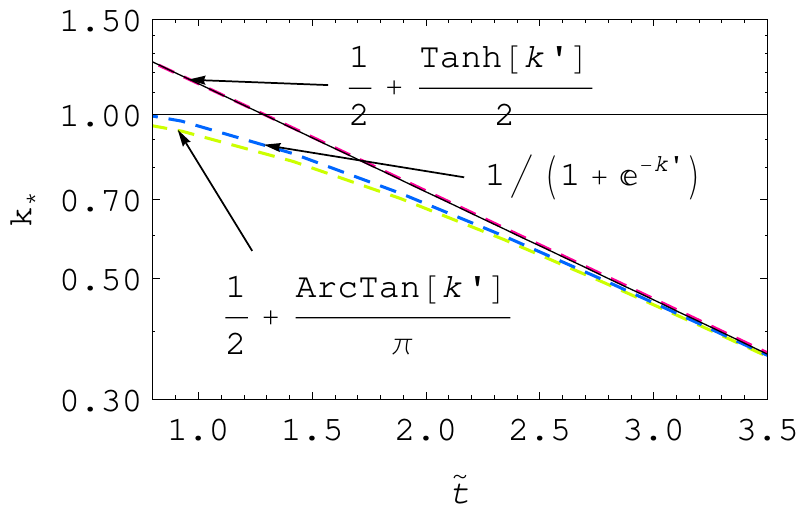}
	\caption{Influence of filter function shape on the comoving transition scale $k_{*}$ for power-law inflation as a function of modified cosmic time $\tilde{t}$ 
			[\cf~\eqref{eq:modifed-time}] for $p = 12$. Filter argument and parameters are fixed as in Fig.~\ref{fig:compare-filter-functions---exponential}.}
	\label{fig:compare-filter-functions---power-law---modified-time-dependence}
\end{figure}

\begin{figure}
	\centering
	\includegraphics[angle=0,scale=1]{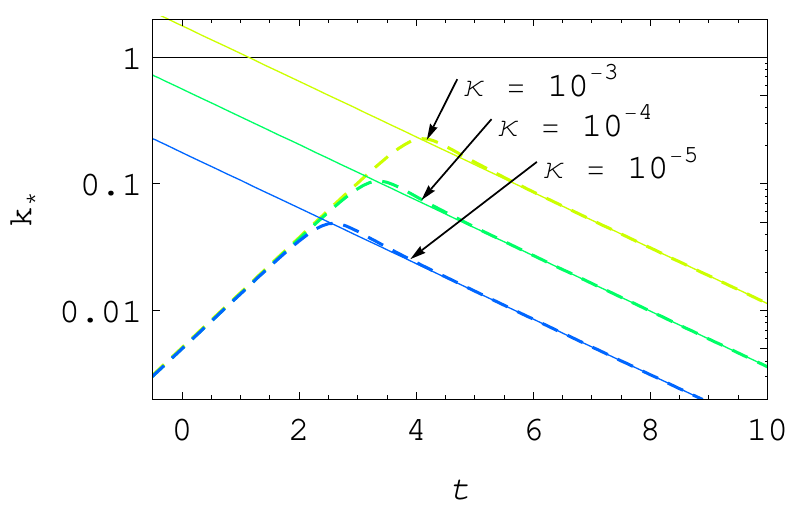}
	\caption{Dependence of the comoving transition scale $k_{*}$ on the width $\kappa$ of filter \eqref{eq:filter-function-main} for exponential inflation 
			as a function of cosmic time $t$ (in units of $H$) for mass $\mu / H = 0.1$ and $\e = 10^{-2}$.}
	\label{fig:compare-different-choices-of-kappa-(exponential)}
\end{figure}

In figure \ref{fig:compare-different-choices-of-kappa-(exponential)} we display the dependence of $k_{*}$ on various values of the width parameter $\kappa$. We find that decreasing this parameter shifts the curves downwards, therefore pushing the dimensional reduction effect to larger and larger scales. This can also be seen directly from the late-time formulae \eqref{eq:kstar-exp}. The power-law case behaves similar.

An observation is that in the limit $\kappa \ra 0$ the dimensional reduction part is absent. This is clear, because there is no mixing of the short quantum modes with the long classical ones in the free (Gaussian) theory as a sharp cut off is introduced. In a slightly different setup, with a filter function having only one parameter, \ie, $\e = \kappa$, this has already been noted in reference \cite{matarrese-2004-0405}.

\subsubsection{Compact Support}
\label{sec:Compact_Support}
We note that all of the filter functions (\ref{eq:filter-1}-f) {\it do not have a lower bound on their support}---a crucial ingredient for the large occurrence of dimensional reduction in the far infra-red. This can be seen as follows: From \eqref{eq:G=G0-sigmaG02} one observes that the dimensional reduction part, $\s\,\Grm_{0}^{2}$, is proportional to $\s$, which itself is related to the filter function $\Wrm_{\kappa}$ in such a way that if $\Wrm_{\kappa} \not\equiv \Theta$ on an interval $\Isf$ only, $\s = 0$ outside of $\Isf$. Consequently, $\Grm = \Grm_{0}$, \ie, dimensional reduction is absent, on the complement on $\Rbb \backslash \Isf$.

However, this does not mean that we can forget about dimensional reduction in the context of stochastic inflation: {\it Any} smooth filter function $\Wrm_{\kappa}$ will definitely cause a deviation from scale invariance (see below), although it might be that this deviation disappears for scales outside the support of $\Wrm_{\kappa}$. Further, the results of this work are not arbitrary, since an ultimate derivation of the stochastic inflation paradigm from first principles will single out a concrete filter function.

We now study the effect of generic filter functions with compact support on the power spectrum. Since $u$ is a free mode function, obeying $( \Box + \mu^{2} ) u = 0$, it follows that the field equation for the long wavelength part does only contain derivatives on $\Wrm_{\kappa}$. For $\Isf = ( - \kappa, + \kappa )$ one may choose
\begin{align}
	\Wrm_{\kappa}( k' )
		&=						\begin{cases}
									0, \q	& k' < - \kappa\\
									k', \q	& k' \in \Isf\\
									1, \q	& k' > + \kappa
								\end{cases} .
\intertext{as a prototype filter function, where $k' \defas ( k \tau - \e ) / \kappa$. Hence, its derivative $\Wrm_{\kappa}'$ is given by}
	\Wrm_{\kappa}'( k' )
		&=						\begin{cases}
									0, \q	& k' < - \kappa\\
									1, \q	& k' \in \Isf\\
									0, \q	& k' > + \kappa
								\end{cases}
\end{align}
leading to the approximation
\begin{align}
	\Wrm_{\kappa}'( k' )
		&=						\begin{cases}
									0, \q															& k' < - \kappa\\
									\exp\!\Big\{\! 1 - \frac{\kappa^{4}}{\left( \kappa^{2} - k'^{2} \right)^{2}} \Big\}, \q	& k' \in \Isf\\
									0, \q															& k' > + \kappa
								\end{cases}.
								\label{eq:W'_kappa_compact_support}
\end{align}

\begin{figure}
	\centering
	\includegraphics[angle=0,scale=1]{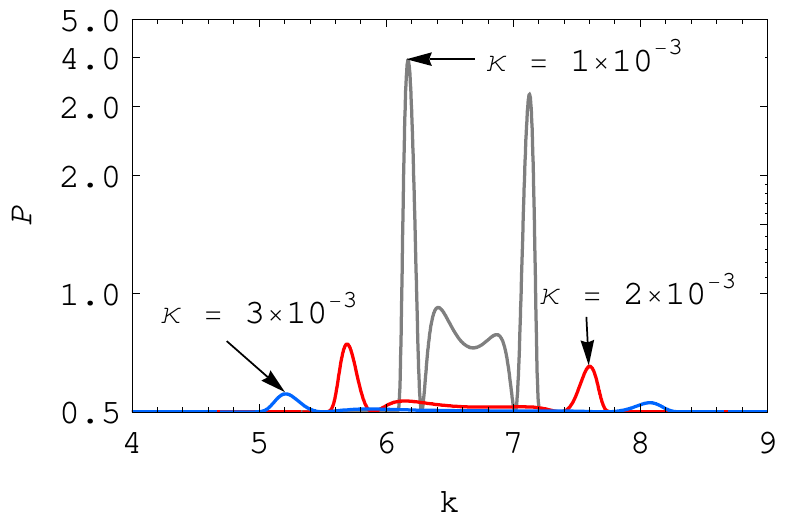}
	\caption{Effective power spectrum $\Pcal( k )$ for filter \eqref{eq:W'_kappa_compact_support} with compact support $\Isf = ( - \kappa, + \kappa )$ 
			for different values of $\kappa$. $\Pcal( k )$ is evaluated for exponential expansion at $t = 6.5 H$ with $\mu = 0$ and $\e = 10^{-2}$.}
	\label{fig:compare-different-choices-of-kappa-(exponential)---finite-support}
\end{figure}

Figure \ref{fig:compare-different-choices-of-kappa-(exponential)---finite-support} shows the influence of $\Isf \ne \Rbb$ on the power spectrum for the filter function due to \eqref{eq:W'_kappa_compact_support}. Although for $k \tau > \e + \kappa$ the dimensional reduction effect disappears, one clearly has an effect inside the interval $\Isf$. As the size of $\Isf$ shrinks, the domain in wave-number space for which the dimensional reduction part is dominant does also size down, albeit the magnitude of the effect of wavelength separation increases considerably. This is reasonable, since the (step-function) limit $\kappa \ra 0$ contains second derivatives on the Heaviside function corresponding to the pole forming for \eqref{eq:W'_kappa_compact_support}.

\begin{figure}
	\centering
	\includegraphics[angle=0,scale=1]{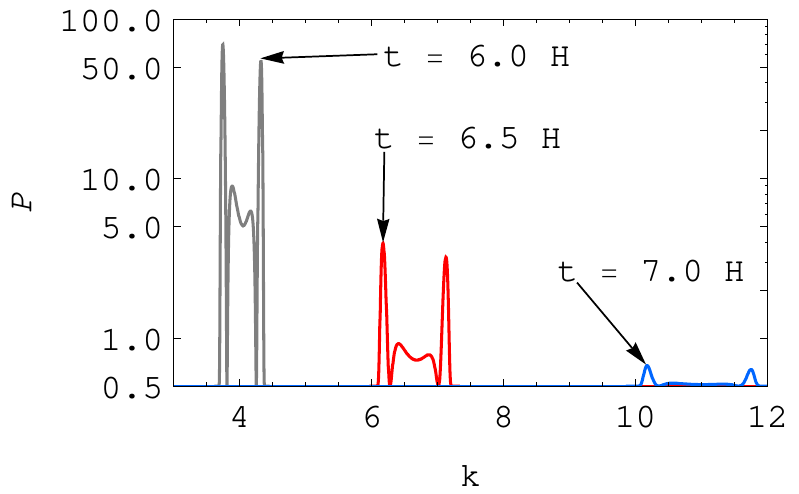}
	\caption{Effective power spectrum $\Pcal( k )$ for various times and $\kappa = 10^{-3}$, otherwise as in 
			Fig.~\ref{fig:compare-different-choices-of-kappa-(exponential)---finite-support}}
	\label{fig:compare-different-choices-of-time-(exponential)---finite-support}
\end{figure}

\begin{figure}
	\centering
	\includegraphics[angle=0,scale=1.25]{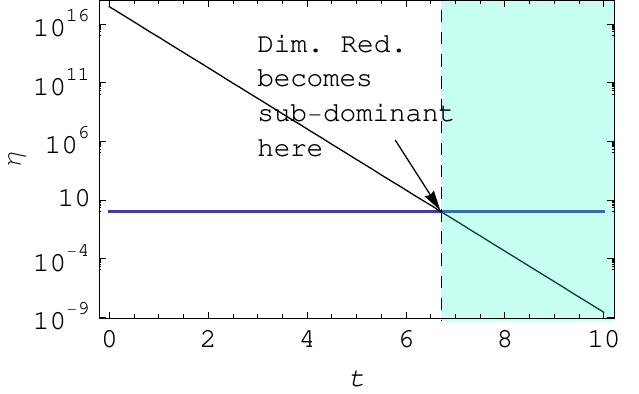}
	\caption{Time dependence of the ratio $\eta$ of the dimensional reduction part to the noiseless part of the effective power spectrum for the filter
			\eqref{eq:W'_kappa_compact_support} for exponential expansion with $\kappa = 10^{-3}$, $\e = 10^{-2}$ and $\mu = 0$.}
	\label{fig:ratio-of-DR-to-free-part-of-the-power-spectrum-exponential---finite-support}
\end{figure}

In figure \ref{fig:compare-different-choices-of-time-(exponential)---finite-support} we display the effective power spectrum $\Pcal( k )$ for different time. As anticipated, the dimensional reduction 'bumps' decline as time increases. One also observes the same behavior as in figure \ref{fig:kstar-exponential-correlation}, namely the grows of the comoving transition scale, up to the point where suddenly the dimensional reduction effect disappears (it is sub-domonant in the shaded region). Plot \ref{fig:ratio-of-DR-to-free-part-of-the-power-spectrum-exponential---finite-support} visualizes the ratio
\begin{align}
	\eta
		&\defas					\frac{\s\,\Grm_{0}^{2}}{\Grm_{0}}
		=						\s\,\Grm_{0}
\end{align}
of the dimensional reduction part to the noiseless part, evaluated at the most ultra-violet peak. One sees that the former diminishes exponentially fast in time. Hence, after some few $e$-foldings, the transition region (from short to long wavelengths) becomes unimportant (shaded region in figure \ref{fig:ratio-of-DR-to-free-part-of-the-power-spectrum-exponential---finite-support}) and the classical power spectrum provides an excellent approximation.

\subsection{Modified Gaussian Fluctuations}
\label{sec:Modified_Gaussian_Fluctuations}
One may connect the replica structure $\s$ to a non-linearity parameter $g_{\t{\tiny NL}}$, which shall now be defined via
\begin{align}
	\varphi^{}_{i}( t, k )
		&\equiv					\varphi^{\t{\tiny G}}_{i}( t, k ) - g_{\t{\tiny NL}}( t, k )\big( \varphi^{\t{\tiny G}}_{i}( t, k ) \big)^{2} ,
\end{align}
where $\vec{\varphi}^{\,\t{\tiny G}}( t, k )$ is a free Gaussian field. On the level of propagators, this translates to
\begin{align}
	{\Grm_{\t{\tiny L}}}( t, k )
		&=						{\Grm_{\t{\tiny L}}}_{0}( t, k ) + 3\,g_{\t{\tiny NL}}( t, k )^{2}\,{\Grm_{\t{\tiny L}}}_{0}( t, k )^{2}
\end{align}
and hence
\begin{align}
	\s( t, k )
		&=						3\,g_{\t{\tiny NL}}( t, k )^{2}
								\label{eq:sigma=13fNL2}
\end{align}
can be directly read off, using equation \eqref{eq:G=G0-sigmaG02}. $g_{\t{\tiny NL}}$ measures the influence of the quantum fluctuations, picked up by a smooth filter function. Formally, it resembles an effective non-Gaussianity parameter \cite{komatsu-2001-63} for the long-wavelength modes. However, this association is misleading since the theory we work with is Gaussian (but with non-trivial replica structure).

\subsubsection{Non-Compact Support}
\label{sec:Non-Compact_Support-gNL}
\begin{figure}
	\centering
	\includegraphics[angle=0,scale=1]{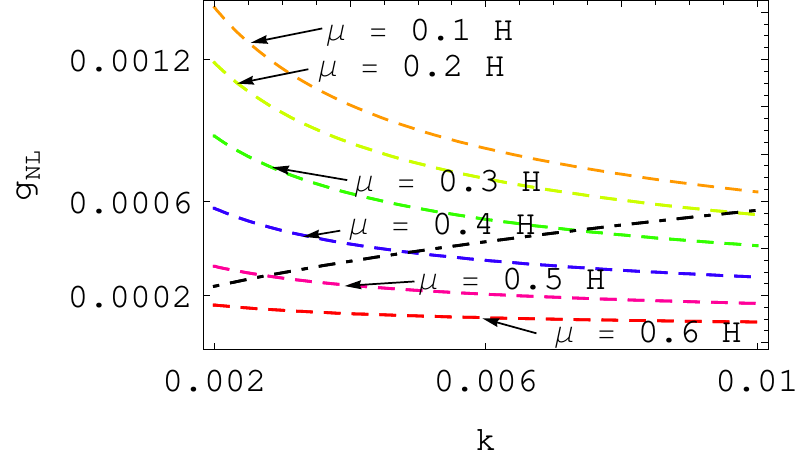}
	\caption{Non-linearity parameter $g_{\t{\tiny NL}}$ for exponential inflation as a function of comoving momentum $k$ (in units of $H$) 
			for mass $\mu / H = 0.1 \t{ (uppermost)},\.0.2,\.0.3,\.0.4,\.0.5$ and $0.6 \t{ (lowermost)}$ at $H t = 10$. 
			For the filter \eqref{eq:filter-function-main} we fix $\kappa = 10^{-3}$ and $\e = 10^{-2}$. The crossings of the dotdashed black line with dashed lines indicates 
			the value of $k_{*}$ for different masses.}
	\label{fig:fNL-exponential}
\end{figure}

\begin{figure}
	\centering
	\includegraphics[angle=0,scale=1]{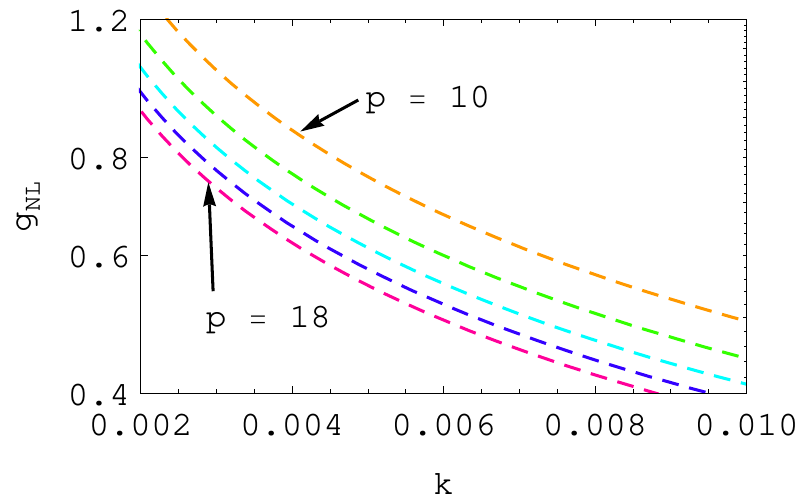}
	\caption{Non-linearity parameter $g_{\t{\tiny NL}}$ for power-law inflation as a function of comoving momentum $k$ (in units of $1 / t_{1}$) for 
			$p = 10\t{ (uppermost)}, 12, 14, 16, 18\t{ (lowermost)}$ and $\tilde{t} = 4$. Filter and parameters are as in figure \ref{fig:fNL-exponential}.}
	\label{fig:fNL-power}
\end{figure}

Let us first consider the case of a non-compact filter function, which we choose to be \eqref{eq:filter-function-main}. Figure \ref{fig:fNL-exponential} shows the dependence of $g_{\t{\tiny NL}}$ on the comoving momentum $k$ for various values of $\mu$ for fixed time $t = 10 H$, using equation \eqref{eq:sigma=13fNL2}. Firstly, one sees that increasing $\mu$ shifts the curve upwards, and secondly, one observers a divergence in the infra-red---displaying the effect of dimensional reduction. For $k \gg k_{*}$ one obtains a scale-invariant spectrum. Figure \ref{fig:fNL-power} visualizes the same with $\mu = 0$ for various $p$, where, $\tilde{t}$ has been fixed to $\tilde{t} = 4$. One observes that increasing $p$ lowers the curves with converge toward their asymptotic value for $p \ra \infty$, which is the same as the $\mu \ra 0$ limit of the exponential case as noted in section \ref{sec:Free_Power_Spectrum}.

\subsubsection{Compact Support}
\label{sec:Compact_Support-gNL}
\begin{figure}
	\centering
	\includegraphics[angle=0,scale=1]{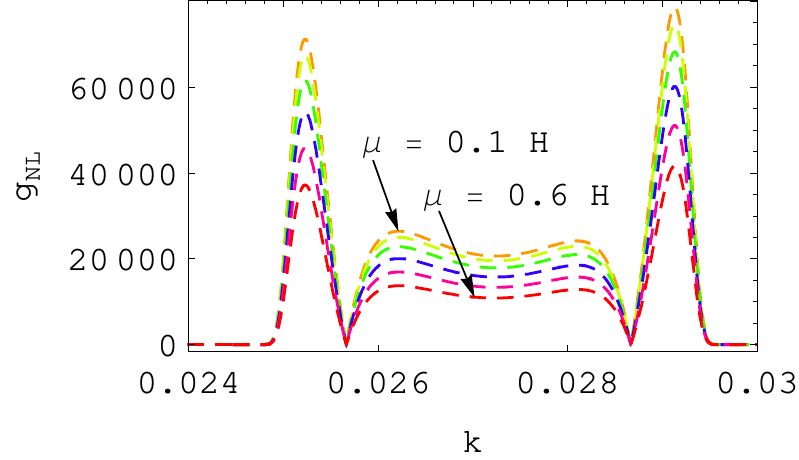}
	\caption{Non-linearity parameter $g_{\t{\tiny NL}}$ as a function of comoving momentum $k$ (in units of $H$) for de Sitter inflation with mass 
			$\mu / H = 0.1 \t{ (uppermost)},\.0.2,\.0.3,\.0.4,\.0.5$ and $0.6 \t{ lowermost)}$ with $H t = 1$. The filter \eqref{eq:W'_kappa_compact_support} with compact 
			support is used with $\kappa = 10^{-3}$ and $\e = 10^{-2}$.}
	\label{fig:g_NL-(exponential)-piecewise}
\end{figure}

\begin{figure}
	\centering
	\includegraphics[angle=0,scale=1]{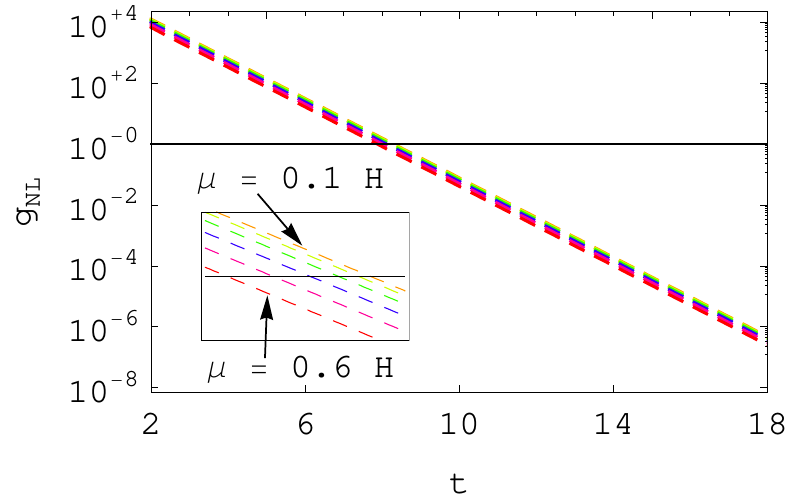}
	\caption{Non-linearity parameter $g_{\t{\tiny NL}}$ as a function of time $t$ (in units of $H$) for de Sitter inflation with mass 
			$\mu / H = 0.1, 0.2, 0.3, 0.4, 0.5, 0.6\t{ (dashed lines, top to bottom)}$ evaluated at the first ultra-violett bump of Fig.~\ref{fig:g_NL-(exponential)-piecewise}. 
			Filter and parameters are as in Fig.~.\ref{fig:g_NL-(exponential)-piecewise}.}
	\label{fig:g_NL-(exponential)-time-dependence-piecewise}
\end{figure}

To discuss the effect of a compact support $\Isf$ on the non-linearity parameter $g_{\t{\tiny NL}}$, we choose the filter corresponding to \eqref{eq:W'_kappa_compact_support}. The dependence of $g_{\t{\tiny NL}}$ on the comoving momentum $k$ is depicted in figure \ref{fig:g_NL-(exponential)-piecewise}. All curves shown in this plot are strictly zero outside the plotted (momentum) intervall. This means that there is only a small ($\sim \kappa$) window around $\e$ in which the wavelengths-separation effects play a role at all. Albeit, for sufficiently early times, the transition effect becomes indeed pronounced. Figure \ref{fig:g_NL-(exponential)-time-dependence-piecewise} shows the time dependence of the non-linearity parameter $g_{\t{\tiny NL}}$ for various masses. After some few $e$-foldings, $g_{\t{\tiny NL}}$ is completely negligible and Gaussianity of the fluctuations in the proper sense holds true.

\section{Summary and Conclusion}
\label{sec:Summary-and-Outlook}
In this work we have studied the large-scale behavior of the power spectrum of the long-wavelength part of $N$-component scalar test fields in curved space-time, using a stochastic description for the quantum modes, introduced by Starobinsky \cite{1982PhLB..117..175S}. We focus on the two important cases of a spatially-flat Friedmann geometry with an exponential and a power-law scale factor. The effective spectral index $n_{\varphi}$ is calculated in the framework of replica field theory {\'a} la Me{\'z}ard and Parisi \cite{1991JPhy1...1..809M}, which we recently introduced in a cosmological context \cite{kuhnel-2008}. Using a Gaussian variational approximation \cite{PhysRev.97.660} we derived an expression for the physical propagator $\Grm$ and thus for the power spectrum of  long-wavelength fluctuations. These methods allow us to study the spatial behavior of arbitrary long-wavelength two-point correlation functions.

A discussion on possible filter functions has been given with special focus on the aspect of the compactness of their support. For filter functions with infinite support we find the phenomenon of dimensional reduction on super-horizon scales. It heavily amplifies the power spectrum in the infra-red. The time-evolution of the long-wavelength field pushes the dimensionally reduced region exponentially fast to unobservable scales. Taking the limit of vanishing width of the filter function $\kappa$, \ie, of a sharp separation of long- and short-wavelength modes, has the same effect.

For filter functions which deviate from the step function on a compact intervall $\Isf$ only, we show that the smooth separation might also lead to strong modifications of the power spectrum. However, this effect is limited to $\Isf$ and decreases exponentially fast in time, becoming negligible after a few $e$-foldings. 

Our findings provide further support for the self-consistency of the idea of inflation. Either regions of broken scale-invariance with extraordinarily large fluctuations disappear faster than any causal patch of the universe expands (non-compact $\Isf$), or large extra power is strongly damped by the time evolution (compact $\Isf$).

The huge effect due to quantum noise on large super-horizon scales (may they occur on a finite or infinite momentum range) do not permit us to speak about the spectrum of fluctuations in the usual, perturbative sense and clearly signal a breakdown of ordinary perturbation theory. This noise modifications further display the failure of the test-field assumption, since in the situation at hand it is no longer valid to neglect the back-reaction of the field on the geometry. In the case of compact $\Isf$ it is possible to avoid the breakdown of perturbation theory and the test field assumption by an appropriate choice of the filter width $\kappa$.

The next step will be to apply our new methods to interacting salar fields and, in turn, discuss replica symmetry breaking solutions. Another open task is to study more general space-times, such as those modelling inhomogeneities.

\section{Acknowledgements}
\label{sec:Acknowledgements}
It is a pleasure to thank J{\'e}r{\^o}me Martin, Aravind Natarajan, Alexei Starobinsky, and Richard Woodard for stimulating discussions. F.K. acknowledges support from the Deutsche Forschungsgemeinschaft (DFG) under grant GRK 881.

\appendix

\section{Noise Distribution and Replica Trick}
\label{sec:Noise_Distributions_and_Replica_Trick}
In this appendix we present some mathematical details and methods needed for our analysis. Because those techniques originate from the study of disordered systems in the framework of statistical field theory, we Wickrotate to an Euclidean action. All formulae can trivially be transformed to those for spaces with signature $(-, +, \ldots, +)$ by simply performing a Wick rotation in the temporal component. Further we generalize from the free case, discussed in the main text, to general stochastic interactions [\cf~\eqref{eq:general-stochastic-interaction-potential}] and work in $d$ space-time dimensions.

First, we explain the replica trick, which provides a simple technique to easily compute stochastic averages of the generating functional, \ie, general $n$-point correlations. Then, we study the noise distribution of the involved random variables, focussing on a Gaussian one. This has the advantage that one can obtain results analytically, being exact in the case of a free theory, though in general only approximative.

\subsection{Replica Trick}
\label{sec:Replica_Trick}
Observables in stochastic systems, \ie, those carrying random variables with a certain probability distribution, adhere---additionally to the classical average over $\varphi_{\t{\tiny L}}$---an average over the noise variables $\hrm$, which shall here be denoted by a bar. In particular, to compute correlation functions and hence power spectra, one needs to compute the noise average of the generating functional of connected $n$-point functions (in the context of statistical physics, the object of interest is the free energy) and thus $\overline{\ln\{ \Zcal[ \vec{\jmath} \,] \}}$, for an external current $\vec{\jmath}$. This is difficult to achieve directly since one has to average a logarithm of a path integral over an exponential. Instead, one uses the so-called {\it replica trick} \cite{1991JPhy1...1..809M}:
\begin{align}
	&\frac{\delta^{n} }{\delta \jmath_{i_{1}}( x_{i_{1}} ) \ldots \delta \jmath_{i_{n}}( x_{i_{n}} )} \overline{ \ln\{ \Zcal[ \vec{\jmath} \,] \} }\notag\displaybreak[1]\\
		&\;=						\lim_{m \rightarrow 0} \frac{m}{m} \frac{\delta^{n} }{\delta \jmath_{i_{1}}( x_{i_{1}} ) \ldots \delta \jmath_{i_{n}}( x_{i_{n}} )}
								\overline{\ln \{ \Zcal[ \vec{\jmath} \,] \} }\notag\displaybreak[1]\\
		&\;=						\lim_{m \rightarrow 0} \frac{1}{m} \frac{\delta^{n} }{\delta \jmath_{i_{1}}( x_{i_{1}} ) \ldots \delta \jmath_{i_{n}}( x_{i_{n}} )}
								\ln\!\left\{ 1 + m\,\overline{\ln\{ \Zcal[ \vec{\jmath} \,] \} \big\} } \right\}\notag\displaybreak[1]\\
		&\;=						\lim_{m \rightarrow 0} \frac{1}{m} \frac{\delta^{n} }{\delta \jmath_{i_{1}}( x_{i_{1}} ) \ldots \delta \jmath_{i_{n}}( x_{i_{n}} )}
								\ln\!\left\{ \overline{\exp\big\{ m\,\ln\{ \Zcal[ \vec{\jmath} \,] \} \big\} } \right\}\notag\displaybreak[1]\\
		&\;=						\lim_{m \rightarrow 0} \frac{1}{m} \frac{\delta^{n} }{\delta \jmath_{i_{1}}( x_{i_{1}} ) \ldots \delta \jmath_{i_{n}}( x_{i_{n}} )}
								\ln\!\left\{ \overline{ \Zcal^{m}[ \vec{\jmath} \,] } \right\} .
								\label{eq:relica-trick}
\end{align}
This means, that one first has to compute $\overline{ \Zcal^{m} }$ for $m$ integer. Then, if the result is an analytic function of $m$, one performs an analytic continuation and takes the limit $m \rightarrow 0$. $\Zcal^{m}$, describes $m$ non-interacting random systems (replicas). By performing the average $\overline{ \Zcal^{m} }$, this $m$ {\it interacting} systems become coupled by means of the stochastic average.

Let us now present, as an application of the replica trick, the computation of the full two-point function.
\begin{widetext}
\begin{align}
	\Grm( x, y )
		&\defas \frac{1}{N}\,\ol{\left\langle \vec{\varphi}( x )\cdot\vec{\varphi}( y ) \right\rangle}
		\simeq					\frac{1}{N}\,\frac{\delta^{2}}{\delta \vec{\jmath}\,( x )\cdot \delta\vec{\jmath}\,( y )}\overline{\ln\big\{ \Zcal[ \vec{\jmath} \,] \big\}}
								\bigg|_{\vec{\jmath}\,\equiv\,\vec{0}}
		\overset{\eqref{eq:relica-trick}}{=}
								\lim_{m \ra 0} \frac{1}{N m} \frac{\delta^{2} }{\delta \vec{\jmath}\,( x )\cdot \delta\vec{\jmath}\,( y )}
								\ln\!\Big\{ \overline{\Zcal^{m}[ \vec{\jmath} \,]} \Big\}
								\bigg|_{\vec{\jmath}\,\equiv\,\vec{0}} \notag\displaybreak[1]\\
		&\phantom{:}=				{\delimitershortfall=-1pt
								\lim_{m \ra 0} \frac{1}{N m} \Bigg[ \frac{1}{\ol{\Zcal^{m}}}\ol{ \int{\!\prod_{a = 1}^{m}\D[ \vec{\varphi}_{a} ]}\,
								\sum_{c,d = 1}^{m} \vec{\varphi}_{c}( x )\cdot\vec{\varphi}_{d}( y )\,\exp\!\bigg\{ - \sum_{b = 1}^{m}\Scal[ \vec{\varphi}_{b} ] \bigg\}}}
								\notag\displaybreak[1]\\
		&\phantom{:= \lim_{m \ra 0} \frac{1}{N m} \Bigg[\;}
								{\delimitershortfall=-1pt
								- \frac{1}{\ol{\Zcal^{m}}^{2}} \ub{\ol{\int{\!\prod_{a = 1}^{m}\D[ \vec{\varphi}_{a} ]}\,
								\sum_{c = 1}^{m} \vec{\varphi}_{c}( x ) \exp\!\bigg\{ - \sum_{b = 1}^{m}\Scal[ \vec{\varphi}_{b} ] \bigg\}}}{\propto}{m}\cdot
								\ub{ \ol{\int{\!\prod_{b = 1}^{m}\D[ \vec{\varphi}_{b} ]}
								\sum_{d = 1}^{m} \vec{\varphi}_{d}( y ) \exp\!\bigg\{ - \sum_{b = 1}^{m}\Scal[ \vec{\varphi}_{b} ] \bigg\}}}{\propto}{m} \Bigg]}
								\notag\displaybreak[1]\\
		&\phantom{:}=				\lim_{m \ra 0} \frac{1}{N m} \frac{1}{\ol{\Zcal^{m}}}\int{\!\prod_{a = 1}^{m}\D[ \vec{\varphi}_{a} ]}\,
								\sum_{c = 1}^{m} \vec{\varphi}_{c}( x )\cdot\vec{\varphi}_{c}( y )\,\ol{\exp\!\bigg\{ - \sum_{b = 1}^{m}\Scal[ \vec{\varphi}_{b} ] \bigg\}}
								\displaybreak[1]\notag\\
		&\phantom{:}=				\lim_{m \ra 0} \frac{1}{m} \sum_{a = 1}^{m} \Grm_{aa}( x, y )
		=						\lim_{m \ra 0} \frac{1}{m} \Tr\Big[ \big( \Grm_{ab}( x, y ) \big) \Big] ,
								\label{eq:Gconderiv}
\end{align}
with $\Zcal \defas \Zcal\big[ \vec{\jmath} = \vec{0}\, \big]$ and
\begin{align}
\begin{split}
	\Grm_{ab}( x, y )
		&\defas					\frac{1}{N\,\ol{\Zcal^{m}}}\int{\!\prod_{c = 1}^{m}\D[ \vec{\varphi}_{c} ]}\,
								\vec{\varphi}_{a}( x )\cdot\vec{\varphi}_{b}( y )\,\ol{\exp\!\bigg\{ - \sum_{d = 1}^{m}\Scal[ \vec{\varphi}_{d} \,] \bigg\}} .
\end{split}
\end{align}
\end{widetext}
In the second step of the first line in \eqref{eq:Gconderiv} we account for the infra-red parts only, where we assume that
\begin{subequations}
\begin{align}
	\ol{\left\langle \vec{\varphi}( x )\right\rangle \cdot \left\langle \vec{\varphi}( y )\right\rangle}
\intertext{has a milder infra-red behavior than}
	\ol{\left\langle \vec{\varphi}( x )\cdot\vec{\varphi}( y ) \right\rangle}
\end{align}
\end{subequations}
\ie, we demand that for $k \ra 0$ the disconnected and the connected correlation functions coincide,
\begin{align}
	\frac{\ol{{\left\langle \vec{\varphi}( x )\right\rangle}\cdot{\left\langle \vec{\varphi}( y )\right\rangle}}}{\ol{\left\langle \vec{\varphi}( x ) \cdot \vec{\varphi}( y ) \right\rangle}}
		&\simeq					0 .
\end{align}
This is exact for the free case discussed in the main text as well as for an even action, \ie, $\Scal[ \vec{\varphi} \,] = \Scal[ - \vec{\varphi} \,]$.

\subsection{Noise Distribution}
\label{sec:Distributions}
Here, we present some general considerations on the distribution for the random variables $\hrm_{\l}$ and introduce some of the statistical physics jargon. For simplicity, we shall assume zero mean, \ie, $\mu_{\l} = 0$. This does not restrict the following discussion, and since we will only average exponentials, we can always shift $\mu$ to zero with the effect of adding a potential, which' coefficients are just given by the corresponding average values, \ie, $\hrm_{\l} \lra \mu_{\l}$. We will see later on that such potentials generate mass corrections. The second cumulant of $\Vrm$ shall then be taken of the form \footnote{Please note the different sign convention for the correlator \eqref{eq:vvbar} compared to the choice of \cite{1991JPhy1...1..809M}.}
\begin{align}
	\overline{\Vrm\!\left( \vec{\varphi}_{a}( x ), x \right)\Vrm\!\left( \vec{\varphi}_{b}( y ), y \right)}
		&=						\phi( x, y ) N \Rrm\!\left( \frac{ \Xi_{ab}( x, y ) }{ N } \right) ,
								\label{eq:vvbar}
\end{align}
where, for later convenience, we rescale by the number $N$ of field components. The function $\Rrm$ reflects the correlation in replica field space, with, \eg,
\begin{subequations}
\begin{align}
	\Xi_{ab}( x, y )
		&\defas					\vec{\varphi}_{a}( x ) \cdot \vec{\varphi}_{b}( y )
								\label{eq:vvbar-times}
\intertext{or}
	\Xi_{ab}( x, y )
		&\defas					\big[ \vec{\varphi}_{a}( x ) - \vec{\varphi}_{b}( y ) \big]^{2} .
								\label{eq:vvbar-difference}
\end{align}
\end{subequations}
We refer to \eqref{eq:vvbar-times} as {\it product correlation}, while \eqref{eq:vvbar-difference} is called {\it difference correlation}. In the following we will mainly use the former (\cf~the discussion at the end of this appendix). Of course, the concrete form of the function $\Rrm$, as well as of its argument have to be determined from first principles. The space-time correlation $\phi( x, y )$ is called {\it short-range}, if $\phi( x, y ) = \wp( t, t' ) \delta^{d - 1}( \xbm - \ybm )$, and {\it long-range} for all other cases (\cf~appendix \ref{sec:Long_Range_Correlation} and also \cite{fedorenko-2007}). Note, that for our example of a free field, the Gaussian distribution is exact and that the function $\Rrm$ is linear. The difference between \eqref{eq:vvbar-times} and \eqref{eq:vvbar-difference} is that the latter has the so-called {\it statistical tilt symmetry}:
\begin{align}
	\vec{\varphi}_{a}( x )
		&\ra						\vec{\varphi}_{a}( x ) + \vec{g}( x ) ,
\end{align}
where $\vec{g}( x )$ is some function without (replica) index.

When performing the noise average over correlation functions, one has to average the measure, and thus, after replication, $\exp\{ - \sum_{a}\Scal[ \vec{\varphi}_{a} ] \}$. Assuming product-correlation and the matrix $( \Arm_{\l,\l'} )$ to be diagonal in components and position (short-range correlation), the part containing noise may be calculated as
\begin{align}
	&\ol{\exp\!{ \left\{ \sumint{\l} \hrm_{\l}\,\varphi_{\l} \ldots \varphi_{\l} \right\} }}\notag\\
		&\q\!\propto				\int{\!\D[ \{ \hrm \} ]}\,\exp\!{ \left\{ \sumint{\l} \hrm_{\l} \varphi_{\l} \ldots \varphi_{\l}
								- \frac{1}{2}\,\sumint{\l, \l'}\hrm_{\l}\Arm_{\l, \l'}\hrm_{\l'} \right\}}\notag\displaybreak[1]\\
		&\q\!\propto				\exp\!\left\{ \frac{1}{2}\,\sumint{\l, \l'} \varphi_{\l} \ldots \varphi_{\l}{\Arm^{-1}}_{\l, \l'}\varphi_{\l} \ldots \varphi_{\l'} \right\}\displaybreak[1]\\
		&\q\!=					\exp\!\left\{ \frac{1}{2}\sum_{a,b = 1}^{m}\int_{x}\!\sum_{j}\Delta_{j}( x )
								\left( \vec{\varphi}_{a}( x ) \cdot \vec{\varphi}_{b}( x ) \right)^{j} \right\}\notag\displaybreak[1]\\
		&\q\!\os{\!}{\equiv}			\exp\!\left\{ \frac{1}{2}\sum_{a,b = 1}^{m}\int_{x}
								N\,\Rrm\!\left( \frac{\vec{\varphi}_{a}( x )\cdot\vec{\varphi}_{b}( x )}{N} \right) \right\} ,\notag
\end{align}
where $\int_{x} \defas \int \d^{d}x$.

This is perhaps a good place to mention a difference between what is described here and what is described in the field of disordered systems. In the latter one studies macroscopic objects, \eg, a crystal, with defects, \eg, vacancies, wrong atoms or molecules, misaligned layers or substrate impurities. These kinds of disorder is mimicked by some random variables - a prime example of such a system is the random field Ising model \cite{PhysRevLett.35.1399}.

Averages over disorder are thought of as averages over different realization, \ie, practically different pieces of a crystal, for instance. M{\'e}zard and Parisi \cite{1991JPhy1...1..809M} made a consistent replica field theoretic approach to those systems, which' $m$ copies, arising due to the application of the replica trick [\cf~\eqref{eq:relica-trick}], are viewed as respectively {\it different}, coming from different realizations, carrying different random variables, with distinct correlations among them.

This is, however, {\it not} the case in stochastic inflation, where the noise arises from the short wavelengths of some quantum field within {\it one and the same} system. For the case under consideration the replica trick \eqref{eq:relica-trick} gives us a simple procedure to perform stochastic averages over the generating functional. There, each $\Zcal[ \vec{\jmath}\, ]$ is a function of the random variables $\{ \hrm \}$, so its averaged $m$'th power, $\overline{\Zcal^{m}[ \vec{\jmath}\, ]}$, is too. It is {\it only} the integration variables $\vec{\varphi}_{\t{\tiny L}}$ that acquires an additional label, namely the replica index $a = 1, \dots, m$.

\section{Variational Method}
\label{sec:Variational_Calculation}
This appendix is devoted to the application of the Feynman-Jensen inequality and a Gaussian variational principle to derive variational equations for a variational propagator $\Grm_{ab}$, which allows us to obtain a closed expression for the power spectrum of the long-wavelength modes.

\subsection{Feynman-Jensen Inequality}
\label{sec:Feynman-Jensen_Equation_and_Gaussian_Variational_Principle}
To perform the stochastic average over the (quantum) noise variables $\{ \hrm \}$, we use the replica trick (\cf~appendix \ref{sec:Replica_Trick}) and obtain
\begin{align}
\begin{split}
	\overline{\Zcal^{m}[ \vec{\jmath} \,]}
		&=						\int{\prod_{a=1}^{m}\!\D[ \vec{\varphi}_{a} ]}\, \overline{ \exp\!{ \left\{ - \sum_{b=1}^{m} \Scal[ \vec{\varphi}_{b}, \vec{\jmath} \,] \right\} } }\\
		&\equiv					\int{\prod_{a=1}^{m}\!\D[ \vec{\varphi}_{a} ]}\, \exp\!{ \left\{\! - {\Scal}^{(m)}\!\big[ \vec{\varphi}, \vec{\jmath} \,\big] \right\} } ,
								\label{eq:Znbar}
\end{split}
\end{align}
with the {\it replicated action} (for short-range product-correlation, for simplicity)
\begin{align}
	&{\Scal}^{(m)}[ \vec{\varphi}\,]
		\defas					\frac{1}{2}\sum_{a = 1}^{m}\int_{t,t'} \int_{\kbm}\Grm_{0}^{-1}( t, t', k )\,\vec{\varphi}_{a}( t, k )\cdot\vec{\varphi}_{a}( t', -k )
								\notag\displaybreak[1]\\
		&\,\hph{=}					- \frac{1}{2}\sum_{a,b = 1}^{m}\int_{x, y}\phi( t, t' )\,\delta( \xbm - \ybm )N\,\Rrm\!\left( \frac{\vec{\varphi}_{a}( x )
								\cdot\vec{\varphi}_{b}( y )}{N} \right) .
								\label{eq:S(m)}
\end{align}
Here and in the following, we refrain from writing the different replica fields in the argument of the action, \ie,
\begin{align}
	[ \vec{\varphi}\, ]
		&\defas					\big[ \vec{\varphi}_{1}, \ldots, \vec{\varphi}_{m} \big]
\end{align}
for simplicity. 

We proceed with a {\it Feynman-Jensen variation principle} \cite{PhysRev.97.660} and therefor define the Gaussian trail action
\begin{equation}
	\Scal_{0}[ \vec{\varphi}\, ]
		=						\frac{1}{2}\sum_{a,b = 1}^{m}\int_{t, t', \kbm}{\Grm^{-1}}_{ab}( t, t', k )\,\vec{\varphi}_{a}( t, k )\cdot\vec{\varphi}_{b}( t',-k ) ,
\end{equation}
where we make the following ansatz for the propagator
\begin{align}
	{\Grm^{-1}}_{ab}
		&\defas					\left( \Grm_{0}^{-1} + \s_{c} + \s_{aa} \right)\delta_{ab} - \s_{ab} .
								\label{eq:Gab-inverse-ansatz}
\end{align}

Let us comment on the structure of \eqref{eq:Gab-inverse-ansatz}. On the main diagonal (in replica space) we have the inverse of the noiseless propagator $\Grm_{0}^{-1}$ {\it plus} some mass correction $\s_{c}$, to be determined later, \eg, by the variational principle described below in this section. This alone would not only be trivial but also inconsistent, as we will see later. Hence, the off-diagonal part is filled by some, a priori unknown, {\it replica structure} $\s_{ab}$, which, in general, can be time dependent, and, if one includes long-range noise correlation (\cf~appendix \ref{sec:Long_Range_Correlation}), also momentum dependent, directly affecting the scaling behavior of the power spectrum. Thus, although this variational method only generates a self-energy contribution, its off-diagonal replica structure might have a viable influence on large-scale correlations. 

The Gaussian variational method becomes exact in the limit $N \ra \infty$ and allows one to go beyond ordinary perturbation theory. It is based on the following Feynman-Jensen inequality \cite{PhysRev.97.660}
\begin{align}
	\ln\{ \Zcal \}
		&\ge						\ln\{ \Zcal_{0} \}
								+ \Big\langle \Scal^{(m)}_{0} - \Scal^{(m)} \Big\rangle^{}_{\!0} ,
								\label{eq:fvar}
\end{align}
where the subscript $0$ refers to the variational action \eqref{eq:S(m)} and we temporarily Wickrotate to Euclidean signature. Equation \eqref{eq:fvar} can easily be proven by using the Jensen inequality $\exp\!\left\{ \langle \ldots \rangle \right\} \le \langle \exp\{ \ldots\} \rangle$ \cite{Jensen-1906-1}, which comes from the convexity of the exponential. The problem is to find the best ${\Grm_{\t{\tiny L}}}_{ab}$, \ie, the best $\s_{ab}$, satisfying (\ref{eq:fvar}) by maximizing the \rhs~of \eqref{eq:fvar}.

Computing $\Fcal_{\t{\tiny var}}$ per component and spatial volume yields
\begin{align}
	&\frac{\Fcal_{\t{\tiny var}}}{N\,\Vol(d-1)}
		=						\frac{1}{2}\sum_{a = 1}^{m}\int_{t, t', \kbm} \Grm_{0}^{-1}( t, t', k )\,\Grm_{aa}( t, t', k )\notag\\
		&\qq\hph{=}				-\, \frac{1}{2}\int_{t, t', \kbm}\Tr\ln\!\big\{ \Grm( t, t', k ) \big\} + C\\
		&\qq\hph{=}				-\, \frac{1}{2}\sum_{a,b = 1}^{m}\int_{t, t'}\phi( t, t' )\,\widehat{\Rrm}\!\left( \int_{\kbm} \Grm_{ab}( t, t', k ) \right) ,\notag
								\label{eq:fvar/NV-times}
\end{align}
where we temporarily switched to finite spatial volume $\Vol(d-1)$. The constant $C$, which vanishes after variation, includes $\big< \Scal_{0} \big>$ as well as terms from $\Fcal_{0}$.

For difference-correlation \eqref{eq:vvbar-difference} one just has to replace the argument of $\widehat{\Rrm}$ to
\begin{align}
	\int_{\kbm}\!\big( \Grm_{aa}( t, t', k ) + \Grm_{bb}( t, t', k ) - 2\,\Grm_{ab}( t, t', k ) \big) .
\end{align}
The ``hat'' over the function $\Rrm$ in \eqref{eq:fvar/NV-times} is defined through
\begin{align}
	\widehat{\Rrm}\big( \!\left\langle \;\cdot\; \right\rangle_{0}\! \big)
		&\defas					\left\langle \Rrm( \;\cdot\; ) \right\rangle_{0} .
\end{align}
In the limit $N \ra \infty$, averaging and applying the (analytic) function $\Rrm$ commute and so we drop the hat when such a limit is considered.

The variation of $\Fcal_{\t{\tiny var}}$ (given in \eqref{eq:fvar/NV-times}) with respect to the $m^{2}$ variational parameters $\Grm_{ab}$ gives for $a \ne b$
\begin{subequations}
\begin{align}
	\s_{ab}( t )
		&=						\phi( t )\,\widehat{\Rrm}'\!\left( \int_{\kbm} \Grm_{ab}( t, k ) \right)
								\label{eq:sc+sab1N-times}
\end{align}
and
\begin{align}
	\s_{c}( t )
		&=						- \phi( t )\,\widehat{\Rrm}'\!\left( \int_{\kbm} \Grm_{aa}( t, k ) \right) ,
								\label{eq:sc+sab2N-times}
\end{align}
\end{subequations}
where we use
\begin{align}
	\frac{\delta \Grm_{ab}( t, \kbm ) }{\delta \Grm_{cd}( r, \pbm )}
		=						\delta( t - r )\,\delta^{(d-1)}( \kbm - \pbm )\,\delta_{ac}\,\delta_{bd}
\end{align}
and define the equal-time entities $s_{ab}( t ) \defas s_{ab}( t, t )$, $s_{c}( t ) \defas s_{c}( t, t )$, $\Grm_{ab}( t, \kbm ) \defas \Grm_{ab}( t, t, \kbm )$, and $\Grm_{0}( t, \kbm ) \defas \Grm_{0}( t, t, \kbm )$.

Again, for difference-correlation we obtain a similar result as \eqref{eq:sc+sab2N-times} (\cf~equation (3.12) in \cite{1991JPhy1...1..809M}) with the argument of $\widehat{\Rrm}'$ replaced by
\begin{subequations}
\begin{align}
	\int_{\kbm}\big( \Grm_{aa}( t, k )
								+ \Grm_{bb}( t, k ) - 2\,\Grm_{ab}( t, k ) \big)
								\label{eq:sc+sab1N-difference}
\end{align}
plus a global minus sign. Additionally, we find in this case
\begin{align}
	\s_{c}( t )
		&=						- \sum_{\substack{a=1\\a\ne b}}^{m}\s_{ab}( t ) .
								\label{eq:sc+sab2N-difference}
\end{align}
\end{subequations}

It is important to note that {\it any} interaction not containing noise just modifies the self energy through its diagonal structure in replica space and that the free Gaussian case studied in the main text, {\it necessarily} yields replica symmetry.

The physical interpretation of the saddle-points equations \eqref{eq:Gab-inverse-ansatz}, (\ref{eq:sc+sab1N-times},b) and (\ref{eq:sc+sab1N-difference},b) is the following: The replica structure $\s$ is a generalized self energy, \ie, the sum of generalized tadpole diagrams (\cf~the discussion in section 3 of \cite{1991JPhy1...1..809M}). This follows from an expansion of the stationarity equations (\ref{eq:sc+sab1N-times},b) and (\ref{eq:sc+sab1N-difference},b) in powers of $(\Grm_{ab})$.

\subsection{Long-Range Correlation}
\label{sec:Long_Range_Correlation}
We now consider the case where the noise-correlation is non-local, or of so-called long-range-type \cite{fedorenko-2007}. This is described by the following correlation \cf~(\ref{eq:vvbar-times},b)]
\begin{align}
\begin{split}
	\overline{\Vrm\!\left( \vec{\varphi}_{a}( x ), x \right)\Vrm\!\left( \vec{\varphi}_{b}( y ),\,y \right)}
		&=						\phi( t, t', \xbm - \ybm )\,N\times\\
		&\hph{=}					\times\Rrm\!\left( \frac{ \Xi_{ab}( x, y ) }{ N } \right) .
								\label{eq:vvbar-times-lr}
\end{split}
\end{align}
Going through analogous steps as in the previous section, we obtain pendants to (\ref{eq:sc+sab1N-times},b) (product correlation),
\begin{subequations}
\begin{align}
	\s_{ab}( t, \pbm )
		&=						\int_{\xbm}\phi( t, \xbm )\,\ee^{- \irm\pbm\cdot\xbm}\,
								\widehat{\Rrm}'\!\left( \int_{\kbm} \ee^{- \irm\kbm\cdot\xbm}\,\Grm_{ab}( t, k ) \right) ,
								\label{eq:sc+sab1N-times-lr}\displaybreak[1]\\
	\s_{c}( t, \pbm )
		&=						- \int_{\xbm}\phi( t, \xbm )\,\ee^{- \irm\pbm\cdot\xbm}\,
								\widehat{\Rrm}'\!\left( \int_{\kbm} \ee^{- \irm\kbm\cdot\xbm}\,\Grm_{aa}( t, k ) \right)
								\notag\\
		&\hph{=}					+ 2\,\widehat{\Urm}'\!\left( \int_{\kbm} \Grm_{aa}( t, k ) \right),
								\label{eq:sc+sab2N-times-lr}
\intertext{and to (\ref{eq:sc+sab1N-difference},b) (difference correlation),}
	\s_{ab}( t, \pbm )
		&=						-2\,\int_{\xbm}\phi( t, \xbm )\,\ee^{- \irm\pbm\cdot\xbm}\,
								\widehat{\Rrm}'\!\left( \int_{\kbm} \ee^{- \irm\kbm\cdot\xbm}\big( \Grm_{aa}( t, k )
								\right.\notag\\
		&\hph{=}					\hs{10mm} \left. \vphantom{\int_{\kbm}}
								+ \Grm_{bb}( t, k ) - 2\,\Grm_{ab}( t, k ) \big)\! \right),
								\label{eq:sc+sab1N-difference-lr}\displaybreak[1]\\
	\s_{c}( t, \pbm )
		&=						- \sum_{\substack{a=1\\a\ne b}}^{m}\s_{ab}( t, \pbm ) ,
								\label{eq:sc+sab2N-difference-lr}
\end{align}
\end{subequations}
where $a \ne b$ for \eqref{eq:sc+sab1N-times-lr} and \eqref{eq:sc+sab1N-difference-lr}. We introduce a non-random potential $\Urm$ (with suitable rescaling by factors of $N$), which is so far taken to be arbitrary. It may arise from non-zero averages of the random distributions (\cf~the discussion at the beginning of appendix \ref{sec:Distributions}). The last equation is restricted to $\hat{\Urm}' \equiv 0$.

Equations (\ref{eq:sc+sab1N-times-lr}-d) already contain our main result, namely the dimensional reduction on large scales. It also shows that the replica matrix is in general space-time dependent, which affects the scaling behavior of the two-point function (c.f.~\cite{PhysRevB.27.5875}). Clearly, the short-range variational equations are included in (\ref{eq:sc+sab1N-times-lr}-d).

\subsection{Replica Symmetric Propagator}
\label{sec:Replica_Symmetric_Propagator}
In the previous section we derived the variational equations (\ref{eq:sc+sab1N-times-lr}-d) for a general matrix $\left( \s_{ab} \right)$. It is important to try the simplest ansatz, which consists of taking $\s_{ab} = \s$ for all $a \ne b$, meaning that different replicas couple all in the same way among each other. This {\it replica symmetric} case is exact for the free case discussed in the main text of this article. One finds
\begin{align}
\begin{split}
	\left( \Grm_{ab} \right)^{-1}
		&=						\left( \Grm_{0}^{-1} + \s_{c} \right)\!\onebbm - \s( \Jbbm - \onebbm )\\
		&=						\left( \Grm_{0}^{-1} + \s_{c} + \s \right)\!\onebbm - \s \Jbbm \\
		&=						\Grm_{0}^{-1}\onebbm - \s \Jbbm .
								\label{eq:Grm-1}
\end{split}
\end{align}
Thus, the inverse $\Grm_{ab}( t, k )$ has the form
\begin{alignat}{2}
	( \Grm_{ab} )( t, k )
		&=						\Grm_{0}( t, k ) \onebbm + \s( t, k )\,\Grm_{0}( t, k )^{2} \Jbbm ,
								\label{eq:grs}
\end{alignat}
where the physical limit $m \ra 0$ has been taken and we define the $m \times m$-matrix $\Jbbm$ by $\Jbbm_{ij} = 1$ for all $i, j$. It has the property $\Jbbm^{2} = m\,\Jbbm$, obtained by inverting \eqref{eq:Grm-1}, is equal to zero in this limit. We observe that the limit of vanishing correlation, \ie, $\s \ra 0$, gives back the free power spectrum as expected and obtained in \cite{matarrese-2004-0405}, with rather different methods.

The {physical} propagator $\Grm( t, k )$ of the long-wavelength field is obtained from $\Grm_{ab}( t, k )$ via [\cf~equation \eqref{eq:Gconderiv}]
\begin{align}
	\Grm( t, k )
		&=						\lim_{m \ra 0} \frac{ 1 }{ m } \Tr\big[ ( \Grm_{ab} )( t, k ) \big] ,
								\label{eq:armean}
\end{align}
which is simply the arithmetic mean of the trace of the replica matrix propagator.

\bibliographystyle{apsrev}

\end{document}